\documentclass[acmsmall]{acmart}

\makeatletter
\def\@ACM@checkaffil{
    \if@ACM@instpresent\else
    \ClassWarningNoLine{\@classname}{No institution present for an affiliation}%
    \fi
    \if@ACM@citypresent\else
    \ClassWarningNoLine{\@classname}{No city present for an affiliation}%
    \fi
    \if@ACM@countrypresent\else
        \ClassWarningNoLine{\@classname}{No country present for an affiliation}%
    \fi
}
\makeatother

\usepackage{amsmath,amssymb,amsfonts}
\usepackage{algorithmic}
\usepackage{graphicx}
\usepackage{textcomp}
\usepackage{multirow}
\usepackage{array}
\usepackage{booktabs}
\usepackage{longtable}  
\usepackage{lipsum}
\usepackage{tabularx}

\newcommand{\cristysbox}[1]{\vspace{0.3em}\setlength{\fboxsep}{0.015\linewidth}\noindent\fbox{\parbox{0.96\linewidth}{\vspace{0.1em}#1}} \vspace{0.3em}}

\usepackage{framed}
\usepackage{changepage}
\usepackage{hyperref}

\newcommand{\interviewquote}[2]{
 \def\FrameCommand{%
    \hspace{0pt}%
    {\color{cyan} \vrule width 2pt}
    \colorbox{white}
  }%
  \MakeFramed{\advance\hsize-\width\FrameRestore}%
  \noindent
  \begin{adjustwidth}{}{1pt}
  {\small``\textit{#1}'' - {#2}}\end{adjustwidth}\endMakeFramed%
}

\newcommand{\chunk}[2]{%
	\fcolorbox{black}{yellow}{\bfseries\sffamily\scriptsize#1}%
   {$\blacktriangleright$#2$\blacktriangleleft$}%
}
\newcommand{\birgit}[1]{\chunk{Birgit}{{\textcolor{cyan}{\textsl{#1}}}}}

\begin{document}

\title{The Factors Influencing Well-Being in Software Engineers: A Cross-Country Mixed-Method Study}

\author{Cristina Martinez Montes}
\orcid{0000-0003-1150-6931}
\affiliation{Chalmers University of Technology and University of Gothenburg, Sweden}
\email{montesc@chalmers.se}

\author{Birgit Penzenstadler}
\orcid{0000-0002-5771-0455}
\affiliation{Chalmers University of Technology and University of Gothenburg, Sweden}
\affiliation{Lappeenranta University of Technology, Finland}
\email{birgitp@chalmers.se}

\author{Robert Feldt}
\orcid{0000-0002-5179-4205}
\affiliation{Chalmers University of Technology and University of Gothenburg, Sweden}
\email{robert.feldt@chalmers.se}

\begin{abstract}

The well-being of software engineers is increasingly under strain due to the high-stress nature of their roles, which involve complex problem-solving, tight deadlines, and the pressures of rapidly evolving technologies. 

Despite increasing recognition of mental health challenges in software engineering, few studies focus on the factors that sustain or undermine well-being. Existing research often overlooks the interaction between personal, collaborative, and organisational influences on this unique population. This study fills this gap by investigating the specific factors affecting the well-being of software engineers. We conducted 15 qualitative interviews and complemented them with a confirmatory cross-country survey to validate and extend our findings to a broader population. Our mixed-methods approach provides a robust framework to identify key factors influencing well-being, including personal perceptions of well-being, interpersonal and collaborative dynamics, workplace support and recognition, organisational culture, and specific stressors inherent to software engineering.

By offering a detailed, context-specific exploration of these factors, our study builds on existing literature and provides actionable insights for improving well-being in software engineering. We conclude with policy recommendations to inform organisational strategies and develop targeted interventions that address the specific challenges of this field, contributing to more sustainable and supportive work environments.

\end{abstract}

\ccsdesc[500]{Software and its engineering → Software creation and management; Software development process management.}
\ccsdesc[500]{Applied computing → Psychology; Health informatics.}

\keywords{Well-being, Human aspects, Resilience, Software engineers}


\maketitle

\section{Introduction} 
\label{sec:intro}

Software development is fundamentally a human activity that relies on engineers' skills, creativity, and well-being. Developers' mental and emotional states significantly impact their productivity and the quality of their work~\cite{graziotin2014happy}. Good levels of well-being enhance cognitive function and job satisfaction, fostering engagement and innovation. In contrast, stress and burnout can lead to decreased performance, more errors, and reduced creativity, ultimately affecting the success of software projects~\cite{penzenstadler2022take}.

Several studies have explored various factors influencing software development, such as personality traits~\cite{cruz2015forty}, feelings~\cite{graziotin2018happens} sentiments and emotions~\cite{cassee2022sentiment} influence software development. Additionally, research has examined the relationship between job satisfaction and perceived productivity~\cite{8851296} and the effects of stress on software engineers~\cite{penzenstadler2022take, montes2024qualifying}. Specific contexts, such as the COVID-19 pandemic, have also been studied to predict well-being and productivity fluctuations under global stressors~\cite{russo2021predictors}.

Despite these contributions, well-being within software engineering remains only partially understood, with significant gaps in how individual, team and organisational factors shape engineers' well-being. 

Well-being can be seen as a dynamic process that allows people to evaluate how their lives progress based on the interaction of their circumstances, activities, and mental resources, often called `mental capital'~\cite{michaelson2024national}. To accurately assess well-being, it is essential to consider both objective factors and personal perceptions.

Wong et al.'s~\cite{wong2023mental} study on mental health is one of the few addressing this gap; however, it focuses on a single country, missing important nuances across different cultural contexts and concentrating primarily on the individual level. This study addresses these gaps by providing a more extended perspective on how intersecting factors influence software engineers' well-being at these three levels.

By having a comprehensive view of the factors influencing software engineers' well-being, we aim to raise awareness about mental health issues in SE and contribute to the literature focusing on the software field. At the same time, we aim to add to the global discussions on improving the workplace. Furthermore, it offers cross-national data on three levels—individual, team, and organisational—allowing for a nuanced understanding of how diverse cultural contexts impact well-being.


With this study, we aim to answer the question:

\textbf{What factors influence the well-being of software engineers?} We wish to understand what in their environment, on a personal as well as a team level, contributes or takes away from software engineers' well-being.

To achieve our goal, we collected 15 interviews and later compared the insights with a cross-country survey, getting 76 valid answers. 
We compare our results to work on wellbeing factors from other fields.


The paper has the following structure: Section II presents the background and related work. Section III explains our mixed-method research, including participant recruitment, data collection, and analysis procedures. Section IV presents the qualitative and quantitative results, and Section V discusses the findings, limitations, and implications for practice. Finally, Section VI introduces future research and concludes our study.

\section{Background and Related Work} \label{sec:related}

\subsection{Background}

This section presents the background and related work for the study at hand. We first present central concepts around well-being and then give an overview of the most relevant related work.

\subsubsection{The Conception of Well-being}

\begin{quote}
    We must practise the things which produce happiness since if that is present we have everything and if it is absent we do everything in order to have it. --- Epicurus
\end{quote}

According to Magyar and Keyes~\cite{magyar2019defining}, the two most common \textbf{lines of well-being research} focus on well-being as a presence of something positive versus an absence of something negative; they have included defining well-being in terms of positive feelings or in terms of positive functioning. The first line of research on \textbf{hedonic} well-being, defined by the degree of positive feelings (e.g., happiness) experienced and by one’s perceptions of his or her life overall (e.g., satisfaction), constitutes and is referred to as emotional well-being, e.g. Diener et al.~\cite{diener2009science}. 
The second line of well-being research is based on \textbf{eudaimonic} well-being, which includes dimensions of positive functioning --- experienced when one realizes their human potential in terms of psychological well-being~\cite{magyar2019defining}.

As framing for well-being in the article at hand, McNaught~\cite{mcnaught2011defining} proposes a \textbf{definitional
framework of well-being}, in which well-being is perceived as a concept concerned with the objective and subjective assessment of well-being as a desirable human state.
In this framework, the central pillars are society, community, family, and the individual.

Further, to frame the relevance of \textbf{intellectual stimulation} for well-being, Anjali and Anand~\cite{anjali2015intellectual} find that intellectually stimulating work increases job contentment and employee commitment in IT. 

Finally, to point out the relevance of \textbf{creativity}, Sokol and Figurska~\cite{sokol2017creativity} confirm creativity as one of the core competencies of knowledge workers, and that it requires space (mentally and on the schedule) to come to fruition. 

\subsection{Related Work}

\subsubsection{General Population}

There are a number of works that investigate well-being in terms of quantitative assessments as well as in specific factors that contribute or limit the perception thereof. 

The most established models were presented by Diener and Seligman: Diener~\cite{diener2009science} looks at individual or subjective well-being and was the first to establish psychometric instruments for measuring the construct, for example the subjective well-being (SWB) scale.
Seligman~\cite{seligman2011flourish}, one of the central figures of positive psychology, establishes a conceptualisation of well-being around the five pillars of positive emotion, engagement, relationships, meaning and accomplishment.

As an often-cited model in popular science and grey literature, Robinson~\cite{rath2010wellbeing} identifies five categories of well-being as essential:
career well-being (how you occupy your time, or liking what you do every day), social well-being (having strong relationships and love in your life), financial well-being (effectively managing your economic life), physical well-being (having good health and enough energy to get things done on a daily basis), and community well-being (the sense of engagement you have with the area where you live).


A more internally focused version is presented by Michaelson et al.~\cite{michaelson2024national}, who propose a framework for personal well-being with five components: emotional well-being, satisfaction with life, vitality, resilience and self-esteem, and positive functioning.
Specifically within an employee context, Nielsen~\cite{nielsen2003work} finds that well-being in self-managing teams depends strongly on supportive management.

Moving more towards the \textbf{limitations} of well-being, Leifels and Zhang~\cite{leifels2023cultural} investigated cultural factors and found that a significant predictor of well-being impairments were lack of trust and accountability in only mono- and bicultural teams, not in multicultural teams. Misunderstanding and disagreement was found to be positively associated with well-being impairments only in multicultural work teams.

More in the direction of \textbf{social utility}, Michaelson et al.~\cite{michaelson2024national} make a case for national governments  directly measuring people’s subjective well-being --- as in: their experiences, feelings and perceptions of how their lives are going ---  to guide societal development. They call for these measures to be collected on a regular, systematic basis and published as National Accounts of Well-being, and argue that the measures are needed because the economic indicators which governments currently rely on tell us little about the relative success or failure of countries in supporting a good life for their citizens. 

In a similar vein, but more oriented towards \textbf{companies}, Harter et al.~\cite{harter2003well} propose to measure the social utility of subjective well-being in terms of business profitability, productivity, and employee retention.

Since there is no work yet on well-being factors in software engineering, we are comparing the factors model that results from our work to the models presented in this subsection.

\subsubsection{Software Engineering Population}
Several important contributions from the last years show the various impacts on well-being from within the software engineering domain.
This includes the state of frustrations of software developers at work~\cite{ford2015exploring}, burnout~\cite{tulili2023burnout}, and (un)happiness of developers~\cite{graziotin2018happens}.
The effects of remote work during the pandemic were investigated both by Russo et al.~\cite{russo2021predictors} and Ralph et al.~\cite{ralph2020pandemic}. 

On the intervention side, Bernardez et al. investigate the mindfulness interventions for conceptual modeling~\cite{bernardez2020effects}, and Penzenstadler et al. conducted studies on the impact of breathwork interventions~\cite{penzenstadler2022take}.

On the \textbf{exploratory and analytic} side, Madampe et al.~\cite{madampeaddressing} investigate reasons for negative emotions in agile contexts and propose several solutions to overcome the causes. One of such negative responses, the experience of feeling overwhelmed, is explored by Michels et al.~\cite{michels} in a qualitative psychology study that identifies seven distinct categories: communication-induced, disturbance-related, organizational, variety, technical, temporal, and positive overwhelm.
De Souza Santos et al.~\cite{Santos} investigate how hybrid work influences the well-being in the software industry. Their findings indicate that hybrid work offers primarily positive effects on the overall well-being, but also has challenges like infrastructure issues and reduced interaction with co-workers. 
Santana et al.~\cite{Santana} identify everyday interpersonal challenges that point to a lack of psychological safety in software development practices, challenges such as reluctance to admit mistakes, avoiding seeking help, and fear of sharing negative feedback.
Leme et al.~\cite{leme2024mental} developed an approach based on the GQM (Goal, Question, Metric) methodology to collect, measure, and monitor metrics associated with mental health and productivity. They found a positive correlation between mental health and productivity.
Finally, Storey et al.~\cite{storey2019towards} developed a theory with a bi-directional relationship between software developer job satisfaction and perceived productivity that identifies what additional social and technical factors, challenges and work context variables influence this relationship. A survey instrument developed to instantiate the theory with a large software company. The results suggest propositions about the impact of various factors as well as challenges on developer satisfaction and perceived productivity. 
In \textbf{contrast}, the study at hand uses \textbf{qualitative data from interviews} to establish well-being factors.

On the \textbf{constructive solutions} side, Dwomoh and Barcomb~\cite{dwomoh} explore three ways in which organizations and individuals interested in improving representation can make tech career more inclusive: by (1) supporting networking, (2) cultivating inclusive leadership, and (3) promoting the development of self-efficacy. 
Cerqueira et al.~\cite{Cerqueira} recommend that team members practice empathy by being mindful, being open, understanding others, and taking care, which can reduce blame, improve job motivation, prevent burnout, and create a better work environment.
Singh et al.~\cite{singh2024softment} worked with women software engineers and provided a prototype that employs emotion detection approach to generate Mental Health Scores, called SOFTMENT (SOFTware sector MENTal well-being support system).

For an \textbf{assessment} perspective, Hicks et al.~\cite{hicks} present a research-based framework for measuring successful environments on software teams for long-term and sustainable socio-cognitive problem-solving that was tested across 1282 full-time developers in 12+ industries; predictive of developers’ self-reported productivity.
Sghaier et al.~\cite{Sghaier} present a conceptual framework designed to assess AI-driven software engineering tasks with the goal of customizing the tools to improve the efficiency, well-being, and psychological functioning of developers.

Most closely related to the article at hand, Wong et al.~\cite{wong2023mental} conduct and analyze 14 interviews with software developers to discuss how mental well-being should be considered within the context of work across individual, team, and organization levels, and highlight the need for integrating mental well-being into the technologies employees use at work. The authors focus mostly on personal experiences with mental well-being in the workplace, along with their approaches to managing it in the US context. 
Our study integrates a \textbf{Europe-centric perspective from the interviews with a global outlook from the survey}, enabling us to uncover broader, cross-cultural patterns related to mental well-being and workplace dynamics.

\section{Methodology} 
\label{sec:methodology}

\subsection{Study design} \label{study design}

This study adopted a mixed-methods approach utilising surveys and interviews to comprehensively explore the factors influencing the well-being of software developers. Interviews were conducted with software engineers working in Sweden, examining the cultural, social, and contextual factors shaping well-being within that specific context. Subsequently, surveys were distributed to software developers across several other countries.

The combination of interviews and surveys allows for a nuanced understanding of the diverse factors contributing to well-being among software developers, both within a specific cultural context and across different cultural settings. Surveys provide quantitative data to analyse trends and comparisons across countries, while interviews offer qualitative insights into the unique experiences and challenges developers face within a particular cultural milieu. 

Inspiration was taken from the Bioecological Model (BM) by Bronfenbrenner \cite{bronfenbrenner2000ecological} to design the data collection instruments and to later analyse the data. The BM, being an ecological approach, embraces holistic views, recognising that biological, psychological, sociocultural, and physical environmental factors collectively influence well-being\cite{stokols2000social}. This approach values both physical and social environments in health creation: physical aspects encompass architecture, geography, and technology within a context, while the social environment includes the cultural, economic, and political dynamics at play\cite{stokols1992establishing}. Hence, the questions in the interview and the survey explored the different systems that the subjects interact with aiming to make connections between personal situations and explain how these intersect with those other systems (team, company and culture).

\subsection{Population}

Our target population was software engineers currently working in IT. For the interviews, we specifically looked for engineers living and working in Sweden. However, we aimed to have software engineers answer the survey from anywhere in the world. We wanted to compare and contrast our results from Sweden to other countries.

\subsection{Data Collection}

We collected data from interviews and surveys. The following subsections elaborate on each instrument and its corresponding pilots and adjustments.\\

\subsubsection{Pilots of the data collection instruments}

We used an interview guide and a survey to collect our data. Both instruments were piloted before applying them to our target population. The interview guide was tested two times to make sure the questions were clear and to measure the estimated time. The first author corrected the guide based on the interviewee's feedback. The survey was piloted at the Eclipse Developer Conference, which took place in Ludwigsburg, Germany, in October 2023, and we got 20 answers. Participants gave feedback on the questions, and corresponding changes were made.\\ 

\subsubsection{Interviews}

Qualitative data was collected through 15 individual semi-structured interviews using an interview guide with open questions to gather in-depth information~\cite{denscombe2009ground}. The interview guide was designed at three levels plus the demographic data. The first questions gathered information about the background and experience of the interviews, the coming sections explored the factors that influence well-being at the individual, team and organisational levels. As mentioned in Section~\ref{study design}, the interview aimed to explore the participants' context in a holistic way considering the systems that the interviewee interacts with. See appendix for interview guide.

To recruit interviewees, we used social media posts such as Linkedin, X, and Facebook groups, direct emails to software companies, and the personal networks of the three authors and the university contacts. We targeted software engineers living and working in Sweden. 

The interviews lasted between 40 and 75 minutes. We gave the participants the option to join online or in person, so we had thirteen interviews in person and two online. The interviews were performed by the first author with the aim of consistency.  The first contact with the participants was to explain the goal of the interview and to share the informed consent. During the interview, the first step was to establish rapport and to present and sign the informed consent which explained the goal of the interview, the voluntary and anonymous participation, and the interviewees' right to withdraw their participation at any time. All interviews were audio recorded with the interviewees' consent, (see informed consent in the appendix) and later transcribed in a denaturalised way (this is removing involuntary vocalisation) focusing only on the content of the interview~\cite{oliver2005constraints}.\\

\subsubsection{Survey}

We designed the survey in a similar way to the interview. The first page of the online survey, showed the informed consent with an explanation of how the data would be handled and let the participants know that participation was voluntary and anonymous. We provided our contact information in case participants had questions or wanted to reach out to us. 

The survey had 33 questions in total. The first questions collected demographic information, while the following sections explored how the perception of well-being, the influence of equality, equity, diversity and inclusion, the relationship with managers and peers, companies' culture and physical environment influence software engineers' well-being. The survey had open, multiple options and Likert scale questions. We tailored the scales based on the questions and answers' options to better capture the participants' perceptions and opinions. For example, the scale for overall well-being is different to the scale to measure how heard and respected the participants feel.

Relevant questions were identified from existing research (e.g.~\cite{wong2023mental}, \cite{diener2009science}) and the preliminary results of the interviews.

The survey was available in three languages, namely English, Spanish and Portuguese. The survey was posted on Linkedin, X, and Facebook. We contacted several software companies to ask for support to share the survey. Similarly, several personalised emails were sent to software engineers inviting them to answer. We targeted software engineers from anywhere in the world.

\subsection{Data Analysis}

\subsubsection{Interview Analysis}
The interviews were transcribed and checked against the original recordings for accuracy. We analysed the transcripts using reflexive thematic analysis following Braun and Clarke's six steps \cite{clarke2021thematic}. After reading the transcript several times to become familiar with the data, the first and second authors coded three interviews (20\% of the total data) to assess coding reliability. We compared our results and for each code, we aligned labels, definitions, and examples. Later, we coded the rest of the transcripts. Then, we continued with the rest of the steps namely, combining codes into themes, reviewing and refining themes and reporting of findings.\\

\subsubsection{Survey Analysis}

The first step was to clean and organise the data. Every survey answer was read to make sure the respondents were among our target group. Answers from people not working in the software field were deleted. Next, the answers in Spanish and Portuguese were translated into English to create one single database for analysis. Based on the type of question, we used graphs to visualise the answers. The demographic data was analysed and summarised to understand participants’ age, gender, area and years of expertise, and geographical distribution.

The Likert scale questions were analysed using descriptive statistics and visually represented with diverging stacked bar charts. The open questions were analysed using content analysis.\\ 


\subsubsection{Reflexivity}

Here, we outline the backgrounds and perspectives of each study author, examining how our unique experiences might have influenced both the research process and its outcomes. This reflexive approach \cite{lenberg2024qualitative} is critical in qualitative research, helping to identify and mitigate biases that might shape the interpretation of findings.

The first author, with a bachelor’s degree in psychology and a master’s in social work, offers a strong foundation in human behaviour and social dynamics, supporting an exploration of well-being factors like stress, coping strategies, and interpersonal relationships within software engineering environments. In contrast, the second author holds a PhD in Software Engineering, paired with training as a yoga instructor and embodied mindfulness coach, which brings a unique balance of technical and mindfulness insights to the study. Their background informs an understanding of work-related challenges, such as workload, deadlines, and technology’s role in daily tasks. The third author, with dual expertise in psychology and software engineering and over two decades of consulting experience, provides an integrative perspective on organizational processes and team dynamics, bridging the human and technical aspects of our research. Together, we share a view that human factors in software engineering are often undervalued and deserve greater attention for creating healthier, more effective organizations.

This blend of interdisciplinary perspectives has shaped our approach. The first author’s insights into psychological and social dynamics, grounded in practical community work, enriched the analysis. The second author’s combined expertise in technical and therapeutic fields contributed to a holistic perspective, integrating rigorous software engineering with mindfulness. Meanwhile, the third author offered a broad organizational view, emphasizing the impact of culture and context as well as individual attributes on well-being.

Throughout the study, the first and second authors led the qualitative analysis with a reflexive stance, regularly evaluating assumptions and biases through open dialogue. The third author acted as an external reviewer, critically examining methodological choices and interpretations. This approach aimed to enhance the study’s credibility and to represent participants’ experiences with integrity. Nevertheless, we acknowledge that our shared belief in the importance of human factors in software engineering may have influenced our interpretations, despite efforts to remain objective.

\subsection{Ethical Considerations}

This research followed the recommendation of ethical research study guidelines of Chalmers University. Further, this study was approved by the Swedish Etikprövningsmyndigheten\footnote{https://etikprovningsmyndigheten.se}. Informed consent was obtained from all participants.

Participants were thoroughly briefed on the study's objectives, methods and potential risks. They were also informed of their right to withdraw from the study at any time without facing any consequences.

To protect participants' privacy, all personal identifying information was kept strictly confidential. Each interview participant was assigned a unique code as an identifier, and all collected data, including transcripts and audio recordings, was anonymised and securely stored. Access to the information was restricted to authorized researchers only.

\section{Results} 
\label{sec:results}

This section presents the results from the interviews and the survey.

\subsection{Interviews}

We conducted 15 interviews with software engineers. Table \ref{tab:demo_data} presents the respondents' positions and years of experience.

\begin{table}
\caption{Demographics of interview respondents}
\centering {\footnotesize
\begin{tabular}{p{7cm} p{4cm}}
\hline
    \textbf{Position} & \textbf{Years of Experience} \\ \hline
    Product Test And Integration Engineer  &  2 \\ 
    Software Developer  & 7 \\ 
    Software Developer  & 5 \\ 
    Software Developer  & 23 \\ 
    Configurations And Test Methods & 20+ \\ 
    Embedded Software Engineering  & 10 \\ 
    Systems Engineer & 7 \\ 
    Software Developer & 5 \\ 
    Software Application Developer & 6 \\ 
    Back-End Developer & 15 \\ 
    Requirements Engineer / Research Project Leader & 6  \\ 
    Scrum Master And Developer & 3.5 \\ 
    Computer Vision Specialist & 7 \\ 
    Software Developer & 12 \\ 
    Software Developer & 12 \\ \hline
  \end{tabular}}
  \label{tab:demo_data}
\end{table}

\begin{figure*}
\centering
    \includegraphics[width=14cm]{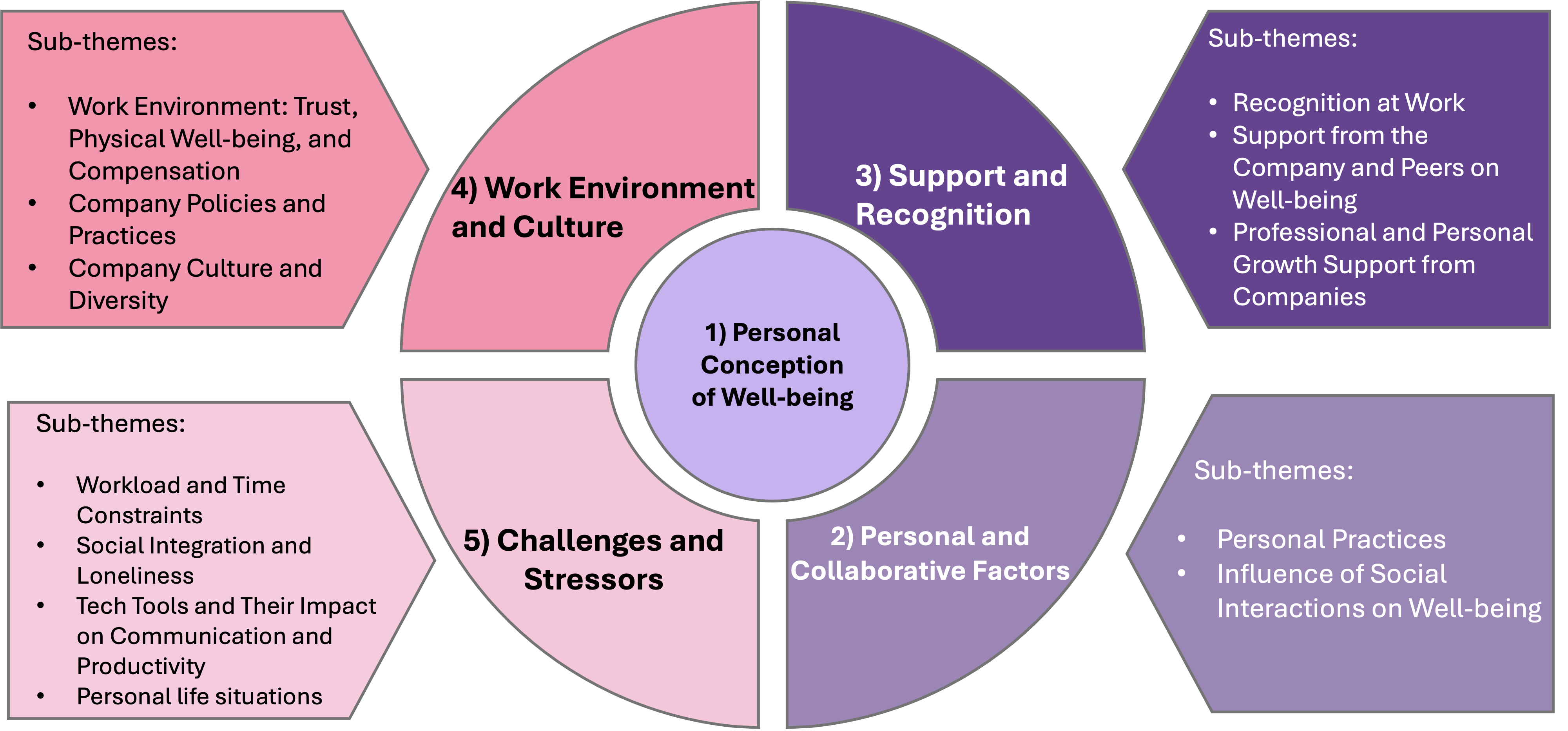}
    \caption{Themes and sub-themes identified in the interviews.}
    \label{themes}
\end{figure*}

From the thematic analysis, five themes emerged. See Figure \ref{themes} for an overview of themes and sub-themes. In the following sections, every theme is explained with its corresponding sub-themes.

\subsubsection{\textbf{Theme 1: Individual Conception of Well-being}}

This theme explains how software engineers conceive their well-being, with most seeing it as a multifaceted concept. 

Well-being for software engineers, according to the interviews, comprises several aspects. It involves feeling happy, content, motivated to perform daily activities, and supported by a healthy work environment. It includes balancing personal and professional life without interference, ensuring mental and physical health, and having safety and access to fundamental human rights. Well-being also encompasses mental and emotional aspects such as the absence of stress and anxiety, sleeping well, not feeling overly tired, and lack of suffering. 

Additionally, it encompasses meaning and accomplishment, including having meaningful tasks, feeling accomplished, and being able to help others. It is about having peace of mind and not being stressed about work deadlines. It comprises physical wellness, which involves feeling physically well, being active, and not getting out of breath easily. Finally, social aspects are crucial, including having supportive relationships and a positive work culture.

In conclusion, software engineers conceive their well-being as a multidimensional concept encompassing emotional, physical, and social aspects. This holistic approach to well-being is reflected in the coming themes.
\\

\subsubsection{\textbf{Theme 2: Personal and Collaborative Factors}}

Starting from a individual point of view, this theme elaborates on the various well-being practices and how SE integrate them into their routines. Physical activity, from gym sessions to yoga, is a prevalent practice. 

Beyond individual practices, and considering the immediate context, social connections significantly influence well-being. Open communication, trust, and mutual respect create positive interactions that foster emotional well-being and reduce stress. Conversely, a lack of support or negative interactions can have detrimental effects. Overall, SE's well-being is shaped by personal efforts and collaborative factors. \\

\paragraph{Sub-theme 1: Personal Practices.}

Several key activities and their regularity were identified. Physical exercise, including gym attendance and sports, is frequently mentioned, with some participants going to the gym three to five times weekly. Yoga and breath work are cited as regular practices. 

The quote below shows the emphasis of one participant on physical activity, although they do not perceive any intentional or specific actions aimed at directly addressing their mental well-being.

\interviewquote{For physical health, I go to the gym but I don't think I'd do anything special for mental well-being.} {p7}

From a different perspective, the participant below mentioned how, for them, physical and mental well-being are connected and taken care of at the same time, viewing physical activity as a foundational aspect of their overall well-being.

\interviewquote{Number one, foremost and having the opportunity to move or maybe I'll keep repeating this over and over again. But taking care of my physical well-being is like one of the best ways I know of taking care of my mental well-being.} {p 10}

The high regularity with which participants engage in physical exercise, three to five times a week, highlights the significance of this practice in their well-being routines. 

Social interaction plays a significant role as a well-being practice, with many respondents regularly going to the office to socialise with colleagues, living with partners, and frequently meeting friends to foster well-being. As mentioned by one participant: 

\interviewquote{I like to come into the office quite often. I can work at home some days, but mostly I want to be at work because I gain something from the social interaction with colleagues.} {p 4} Social engagement plays a crucial role in enhancing mood and overall well-being.

Participants also mentioned several activities they practice regularly, such as taking walks, yoga, breathwork (for example box breathing), hobbies, meditation, and positive affirmations, showing a holistic approach to well-being. 

One participant mentioned when asked what they do as well-being practice:
\interviewquote{Not really. Nothing specific at least for that purpose, other than general you know, hobbies and everything, nothing specific for well-being} {p 3} 
While they may not engage in specific practices targeted at well-being, by incorporating activities that bring joy and fulfilment into their lives, individuals enhance their psychological resilience and cultivate a sense of work-life balance.

Additional activities include going to the mall, sleeping well, meditation and acro-yoga, singing in a choir, Cognitive Behavioral Therapy (CBT), participation in marathons and seeing a psychologist has been a regular practice for other participants for a few years. 

\interviewquote{Being active is like, I feel like I get more dopamine when I am more active, including going to the mall, maybe going for a walk. And also, as I said, hanging out with friends and just going out instead of staying indoors.} {p7}

These findings show how participants integrate physical, mental, and social activities with varying regularity. It also gives an idea of the different angles of well-being. Participants tailored their activities to their individual preferences and needs.\\

\paragraph{Sub-theme 2: Influence of Social Interactions on Well-being}

Participants reflected on how their well-being is influenced relationships, and interactions with others. Participants see social connections, both at work and in personal life, as crucial for well-being. The interactions characterised by open communication, trust, and mutual support, provided emotional support and a sense of belonging. Conversely, challenges such as communication barriers or conflicts, can create stress and negatively affect individuals' mental health. 

One participant mentioned how impactful it is for them to have friendly colleagues, highlighting at the same time the role of positive social interactions in the workplace, where friendly relationships contribute to an individual's emotional well-being and overall satisfaction at work.

\interviewquote{I love having friends. Like hanging out with people that I like... So it is important for me, having friendly colleagues, that I can talk to them freely.} {p7}

Participants mentioned activities that promote emotional sharing and help resolve conflicts, enhancing team cohesion and reducing stress. For instance, open communication seemed to flow better during team events and after work. Furthermore, personal relationships outside work also play a crucial role in our participants' well-being. One participant mentioned about the importance of the people around them:

\interviewquote{So I would say the people around me really matters for me. So, if they're bringing negative vibes, it really affects me. the people is the main factor that makes me feel mentally well. So, if I feel alone or if I feel you know, left out, I definitely feel down and I'm sad.} {p14}

This quote shows how positive and negative personal relationships can foster positive and negative, and influence one's overall sense of well-being, which, in turn, influences work performance and satisfaction.

In conclusion, this sub-theme showed that the well-being of software engineers is significantly influenced by their social interactions and personal relationships characterised by open communication and mutual respect. Conversely, negative interactions and lack of support can increase stress and decrease general well-being. \\

\cristysbox{
\textbf{Theme 2 takeaway:} Software engineers achieve well-being through personal practices and social connections. Regular physical activities are essential to physical and mental health. Additionally, positive social interactions enhance emotional well-being, while negative or unsupportive relationships increase stress.
}



\subsubsection{\textbf{Theme 3: Support and Recognition}}

Two aspects in the work environment, were mentioned by participants to play a key role in their well-being, support and recognition. This theme elaborates on them and explains how respondents perceive their company to provide support through team collaboration, managerial assistance, and access to resources. Additionally, whether recognition is present or not. 

\paragraph{Sub-theme 1: Recognition at Work}

Participants mentioned recognition at work as a factor influencing their well-being and job satisfaction. They elaborated on what recognition at work entails for them and its significance. They also stressed the need for positive feedback and the sense of being part of a team. Recognition, for our participants, involves acknowledging hard work and achievements, providing feedback, and ensuring employees feel integrated. Feeling valued and acknowledged for contributions can significantly enhance motivation and engagement.

Conversely, the absence of recognition can lead to dissatisfaction and even consideration of leaving the job. Interviewees shared their varying experiences and perceptions regarding recognition at their workplace. One of them mentioned: 

\interviewquote{I think I've earned my way into people, at least into my company, and into my peers, and I feel everyone respects me and listens to me when I have something to say.}  {p 1}

This interviewee felt valued for their contributions and believes they have earned the respect of their colleagues and peers. The quote shows that recognition is not only about formal acknowledgements but also about everyday interactions where one's input is valued.

In contrast, other participants mentioned the absence of recognition and how it made them feel. The participant below mentioned that the lack of acknowledgement for their efforts is preventing them from being fully happy at work.

\interviewquote{Well, I'm missing the recognition. That would make me fully happy.} {p 13}

This quote highlights recognition's critical role in an employee's emotional well-being, job satisfaction and career decisions. Further, it shows that without it, even other positive aspects of the job may not suffice to ensure complete job satisfaction. The interviewee mentioned that they were considering changing their job since they still needed the recognition part. Recognition at work is something to consider when planning actions to can influence retention and engagement.\\

\paragraph{Sub-theme 2: Support from the Company and Peers on Well-being}

Support from the company can manifest through various initiatives aimed at promoting mental and physical health and fostering a positive and inclusive work environment. It is common for companies in Sweden to provide allowances catering to mental and physical health. Employees can choose activities that help them manage stress and maintain a healthy work-life balance, as the quote below shows:

\interviewquote{If you want to, they have these programs you can participate in different activities. So if you're interested in a sport, you can participate in clubs. But I mean, it's nothing that you know, unless you look for it or went in the portal search for it. But it's there.} {p 2}

Companies support various activities; however, the employees are the ones who take proactive steps to maintain their overall health. Nevertheless, the effectiveness and perception of this support vary among employees, as the quote below shows.

\interviewquote{My first first thing that I want to say is that there isn't much support from them. Apart from what is in the collective agreement that they need to provide this free sports (wellness allowance) and things like that, which I think is just the bare minimum. They do the bare minimum.} {p10}

This contrasting quote provides a critical perspective, indicating that not all employees feel adequately supported by their company. The respondent perceives the company's efforts are limited to the minimum requirements stipulated by collective agreements without going beyond to offer something meaningful. 

While some employees feel that support is minimal and meets basic requirements, others appreciate different forms of support their companies provide, such as creating a positive and engaging work environment. Participants commented that their companies focus on teambuilding activities and cultural events to strengthen interpersonal relationships and foster a sense of community among employees.
   
\interviewquote{Yeah, so company's trying to be in the best workplaces in the industry in the city. So they're promoting let us... teambuilding and, you know, a lot of cultural balance events every month, so they have trying to have a positive work environment for everyone.} {p14}

This quote illustrates the company's actions to create an engaging and supportive workplace culture. Initiatives such as regular events aimed at cultural balance can foster inclusive work environments and promote that employees feel valued and included. 

In addition to company-led initiatives, peer support adds to the collaborative and positive work environment by creating and fostering an environment where team members can rely on each other for assistance, feedback, and camaraderie. Interviewees elaborate on the impact of their peer network.

\interviewquote{Oh, I don't have anything negative to say because our team is really friendly and we can talk to each other without any hesitation. They all are reachable, even though people are not working in the same office.} {p7}

This participant commented on the importance of open communication and accessibility among team members. The respondent highlights the friendliness and approachability of their peers, which creates an environment where individuals feel comfortable sharing their thoughts and seeking help. It is notable, too, that the quote mentions that the approachability applies to even team members who work in a different location, so peers feel supported regardless of their location.

More positive attributes of the teams were mentioned, highlighting friendliness, supportiveness, and reasonableness. Interviewees, in general, commented on how impactful peer support is on the individual's overall well-being and job satisfaction. \\

\paragraph{Sub-theme 3: Professional and Personal Growth Support from Companies}

This sub-theme focuses on opportunities and support provided by the company for employees to develop professionally and personally. Participants' view of their companies' support showed a complex picture. On one hand, interviewees expressed a potential disconnection between their desires for more growth opportunities and the current company offerings. 

\interviewquote{I don't feel my company supports so much the personal development and the professional development. But I would like it to. I would like to be part of a company that talks more about personal development and professional development. Right now I don't feel it.} {p12}

On the other hand, some participants perceive support from their companies via efforts to provide opportunities for learning and development through platforms and goal-setting. The quote below is an example of that perception.

\interviewquote{The company invest on us, like for our day to day learnings. They have different platforms to learn and there is a platform we can go and learn from there and do the examination and improve our qualifications. Also they have this yearly milestone plannings for the each employee so that they review them by every six months.} {p14}

This interviewee sees the company's provision of learning platforms as a significant factor in professional growth. They also value the company's investment in resources that enable them to learn and stay updated in their field continuously.

In the cases where the companies were not supportive, participants commented on some managers' significant role in taking initiative in employee growth despite the lack of a structured system. One participant mentioned their manager actively supports personal and professional development through regular meetings.

\interviewquote{My manager is actually a really busy person when I look at his calendar, it's always full. But still he finds his time to talk to each each of us. Like we have, like official one on one meetings every two weeks. Other than that, still he talks to us even though he is not involved in what we're doing. He tries to talk to us and see if we face any issues and like not micromanagement, but he is so supportive.} {p 7}

This quote highlights the potential impact of good leadership and a personalised approach to growth. Further, some aspects of the work environment might indirectly contribute to growth, even if not explicitly designed for it. For example, a supportive manager with open communication and a focus on work-life balance can create a positive environment for learning and development. Similarly, opportunities for interaction and support within the team can foster knowledge sharing and a sense of community, which can contribute to personal and professional growth. One participant shared about their colleagues:

\interviewquote{They're always supportive people. They are always helpful. When you ask someone for help you get your help. I always get help from people.} {p15}

While some companies might not have a robust growth support system, there are lines of support from some managers and colleagues that can contribute to employee growth. \\

\cristysbox{
\textbf{Theme 3 takeaway:} Support and recognition are essential for employee well-being and satisfaction. Recognition, both formal and informal, boosts motivation, while its absence may lead to dissatisfaction. Peer, managerial, and company support enhance well-being through mental, physical, and professional growth initiatives.
}


\subsubsection{\textbf{Theme 4: Work Environment and Culture}}

This theme explores higher levels in the BM, focusing in the work environment and culture of the participants' company.\\

\paragraph{Sub-theme 1: Work Environment: Trust, Physical Well-being, and Compensation}

Several participants mentioned trust as an essential aspect they find and want to keep in their work environment. They mentioned that they are more likely to thrive and contribute positively when they feel trusted. One interviewee emphasised the importance of feeling trusted and having flexibility stating: 

\interviewquote{I don't think I would thrive in an environment where they tell me - you need to work from eight in the morning to five in the afternoon every day. Because things happen in life and sometimes you need to be a bit more flexible. So for me, that's really important, flexibility and the trust that comes with that flexibility.}	{p1}

Moreover, trust extended beyond mere sentiment for the interviewees, reflected in management's actions and policies. They pointed out that when upper management conveys a sense of trust in their abilities, it permeates the organisation. As one employee noted:

\interviewquote{They (managers) promote this hybrid work, so we have to go two days a week, even that's not necessary, it's recommended, and they do have the trust, and you feel that they don't micromanage you, you have your own, partially at least, freedom to do yourself. Yeah, really positive culture for sure.}	{p9}

By trusting participants to manage their own time and tasks effectively without needing constant oversight or micromanagement, the company cultivates a flexible and autonomous work environment, fostering a sense of empowerment and accountability. 

One more important aspect mentioned by interviewees was the physical work environment. They commented on the physical well-being tied to the physical workspace including ergonomic and standing desks, chairs and natural light. One interviewee expressed how important it is to consider several factors to create a conducive and comfortable workspace.

\interviewquote{We have nice desks and nice chairs and things like that. The desks raise and lower but the general open office area is catastrophic. It's bad light, we don't get any daylight at all.} {p10}

This quote illustrates a disparity between the physical comforts provided by the office, such as nice desks and chairs with adjustable heights, and the overall ambience of the workspace, particularly the open office area. Despite ergonomic furniture, the environment is described as "catastrophic," primarily due to poor lighting and the absence of natural daylight.

Finally, the salary and benefits were also considered crucial during the interviews. Several participants expressed contentment with their compensation and benefits, not necessarily because it is high but more due to being happy with other company factors such as the work environment. A few commented that their salaries need improvement, such as salary transparency and equitable distribution of benefits across job levels.\\

\paragraph{Sub-theme 2: Company Policies and Practices}

This sub-theme presents diverse participants' perspectives on company policies and practices, highlighting how these influence their experiences, well-being, and organisational engagement. Interviewees expressed value for well-being programs and initiatives provided by the company, such as wellness allowance, lunch walks, and opportunities for physical activity. Conversely, some others expressed dissatisfaction with the adequacy of these initiatives, suggesting the need for more comprehensive well-being support.

\interviewquote{We have asked for higher wellness allowance. The company says no, we will not increase it even thought that benefit make the employee feel better or exercise more. They don't promote any well-being efforts or activities. It feels that the company wants to pull in in every different cost. I don't think they mind if someone, for instance, sent out and hit the wall. It's not like we have any active prevention of being too stressed, Sadly, I'm missing that.} {p12}

This quote shows how some interviewees think the company could do more to prevent stress and promote mental well-being. This feeling was shared by several participants, concluding that the provision of wellness programs and health-related benefits are insignificant.

Another essential aspect mentioned by participants was effective communication and collaboration, which were pointed out as crucial components of company culture. Further, participants said they value open dialogue, feedback mechanisms, and teamwork and peer support opportunities. One participant highlighted the importance of these elements by saying:

\interviewquote{I feel like they're supporting it by giving quite a lot of room to express my opinions and also be able to affect how we do things.} {p4}

The quote shows the importance of a work environment where employees feel heard and empowered to contribute to decision-making processes.
Other participants noted the significance of a collaborative atmosphere, indicating that a supportive culture is vital for personal and professional growth. Moreover, structured team events and informal practices such as open-door policies and peer support were commented to play a significant role in fostering a collaborative environment. One employee mentioned: 

\interviewquote{They tried always to make this mix. To make the people communicate with each other. They remind people in meetings to talk to each other.} {p15}

In conclusion, effective communication and collaboration add to a positive work culture. Participants see it as crucial to foster practices that make them know their opinions are considered, feedback is constructive, and there are ample opportunities for teamwork and support.  \\

\paragraph{Sub-theme 3: Company Culture and Diversity}
Participants commented on various aspects of company culture regarding diversity, touching upon openness to different races, genders, and backgrounds and efforts towards inclusion and equal opportunities. They shared observations and experiences regarding diversity initiatives, policies, the composition of teams, the impact of cultural diversity on workplace dynamics and societal norms regarding diversity and inclusion. While some saw progress and positive steps towards inclusion, others highlighted challenges such as gender imbalances and the persistence of glass ceilings. 

One recurring aspect was the participants' acknowledgement of efforts made by their companies to embrace diversity, such as actively recruiting employees from various backgrounds and promoting inclusivity in hiring practices. 

\interviewquote{We have a lot of employees from different parts of the world, different countries. And we also work with people from other countries.} {p4}

Another aspect expressed in the interviews was the impact of cultural diversity on workplace dynamics. Participants commented their opinions on working in multicultural teams and the value they see in having colleagues from different backgrounds. They recognised that diversity brings different perspectives, enriching discussions and problem-solving processes. However, they also acknowledged the challenges that can arise, such as language barriers or cultural differences in communication styles. Despite these challenges, many believed in the importance of diversity and its positive impact on team dynamics and overall organisational culture.

Moreover, interviewees commented on company policies and practices in shaping diversity initiatives. While some employees perceived their companies as actively promoting diversity through recruitment strategies and inclusive policies, others expressed scepticism about the effectiveness of these efforts, and others mentioned they do not mind diversity in their workplace. 

One participant shared their experience as a minority and how intersectionality plays a role in broadening the issue of diversity.

\interviewquote{So I work in the aviation sector, and that's very male-dominated, very old male-dominated, so it's not so it's not only a sex it's also an age.} {p1}

This quote exemplifies well the challenges faced in industries with entrenched gender and age biases. More participants also shared stories of feeling alienated or marginalised due to their background, while others expressed gratitude for working in environments where diversity is celebrated. One story is the quote below:

\interviewquote{We have really good diverse teams, and about inclusion. Let me tell you one thing. One, day three of my Swedish colleagues were talking to each other and I was there, I was not actively involved in that conversation but these three were talking in English. So I just asked, why are you speaking in English? You can speak in Swedish. So they said, because you're besides us. And if you feel like joining our conversation you can join, if we talk in Swedish then you don't understand. So we have that kind of culture.} {p7}

This quote highlights how crucial it is to create spaces where individuals from all backgrounds feel valued and included and have the opportunity to integrate into their workplace. \\

\cristysbox{
    \textbf{Theme 4 takeaway:} A supportive work environment and inclusive culture significantly impact employee well-being, engagement, and retention. Participants value trust, flexibility, quality workspace, fair compensation, and policies that encourage open communication, collaboration, and wellness. Diversity efforts are appreciated for enriching teamwork. However, challenges like language barriers, gender imbalances and biases still persist.
} 



\subsubsection{\textbf{Theme 5: Challenges and Stressors}}
This theme focuses on the different challenges and factors that motivate stress in our participants.\\


\paragraph{Sub-theme 1: Workload and Time Constraints}

Various factors, including deadlines, customer demands, and the allocation of responsibilities, influence the workload of our participants. They commented that the pressure on them to perform escalates due to the organisation trying to meet delivery targets, particularly when client expectations clash with the organisation's internal capacity. The lack of proper planning leads to a backlog of tasks and increased stress among the interviewees. 

For the interviewees, having a sense of control over their workload is important since it gives them the feeling of handling responsibilities without feeling overwhelmed by stress. However, they also commented that an overload of tasks and unhappy clients can bring down their motivation and make it hard to get things done. In busy times, organised workplaces provide relief. \textbf{Good planning, structured work environments and support from managers} were mentioned as facilitators of handling workload and avoiding feeling overwhelmed.

\interviewquote{One thing that I have seen that the company, or at least the department, has done that is quite negative in my point of view is that there have been people agreeing on deliveries with customers while not having first checked that we have the capacity to fulfil that.} {p5} 

This quote expresses the discussion of workload dynamics, the pressure to meet delivery targets, and the consequences of a lack of proper planning.

\interviewquote{When they get frustrated and when people leave. When I started, one guy had just quit without having a new job. Just he needed to get away. It was horrible, apparently. We have that still to some extent, the frustration within the organization can be, the levels can be high.}  {6}

\paragraph{Sub-theme 2: Social Integration and Loneliness}

One important aspect that directly influences interviewees' well-being was their social integration and feelings of loneliness and exclusion. Participants expressed that they face challenges when integrating socially into their teamwork and making friends. Several of them have struggled to feel included and build meaningful connections. Feelings of shyness, difficulty initiating conversations, and the absence of a close-knit social circle contribute to loneliness and isolation. Despite being immersed in work environments, participants expressed a longing for deeper connections beyond professional interactions. 

\interviewquote{I do have these problems with finding the right people, like, the right friends.}  {p15}

As expressed in this quote, some participants deal with a sense of loneliness at the workplace and in their private lives.  \\

\paragraph{Sub-theme 3: Tech Tools and Their Impact on Communication and Productivity}

The role of tech tools was mentioned as another factor that can lead to stress, frustration, and delays among interviewees. They commented that they face issues with tools that crash and slow IT department responses. One participant noted: 

\interviewquote{We have tools that crash a lot, and the IT department needs to be involved because they are so slow. I have software that I need now to do one specific job within one project, and it’s the 3rd week, and it took, I don’t know how long, it’s a standard software that is available on the web, and it took forever to get access to it.}   {p6}

This participant highlighted the recurring frustration of dealing with unreliable technology. Such problems slow the workflow and cause a ripple effect on project timelines. Furthermore, interviewees also mentioned that restrictive IT policies and outdated tools further hinder productivity, making routine tasks unnecessarily cumbersome and time-consuming. 

Another reason mentioned was the inefficiencies in workplace communication, such as unnecessary meetings, that tools like Zoom or Teams promote that could be replaced by emails. Some participants commented that they felt frustrated and preferred face-to-face interactions over virtual meetings for collaboration.\\

\paragraph{Sub-theme 4: Personal life Situations}

Interviewees explained the main factors from their personal life that influenced their work performance and overall well-being at work. One primary concern was that managing personal responsibilities, such as family issues and tasks, added to the stress burden, making it difficult to maintain a healthy work-life balance. Some participants deal with specific situations, such as conditions like ADHD. 

One participant shared a scenario when they had to deal with different responsibilities at the same time and how they perceived it affected their mental health.

\interviewquote{When we have a lot to do at work and also personally, when there are things I need to take care of, help someone, family, something like that,  it can be anything. Sometimes it can be stressful and it affects our mental health.}  {p2}

Factors such as sleep quality, health issues, seasonal effects like reduced daylight hours in winter, and time spent in social media were mentioned as influencing work performance and negatively impacting mental health. Furthermore, participants commented that the physical environment and daily routines, such as lengthy commutes, also contribute to stress levels.

Another factor mentioned by participants was \textbf{financial pressures}; with inflation rising, managing financial responsibilities, such as mortgages, has become increasingly challenging. On the professional front, interviewees expressed \textbf{feelings of inadequacy and pressure} exacerbated by working alongside highly talented colleagues. They noted that a competitive environment can lead to self-doubt and increased stress as they strive to match the performance of their peers. Finally, one participant commented on the \textbf{agile way of work}; for those who like structure and clear responsibilities, working in agile negatively impacts their well-being.

\cristysbox{
    \textbf{Theme 5 takeaway:} Participants face multiple pressures, including workload, social integration, technology issues, and personal life demands (e.g. family responsibilities and financial pressures), which collectively impact their well-being and job satisfaction. Additionally, the work environment's competitive nature and agile workflows can exacerbate feelings of inadequacy and add to participants' stress.
} 

\subsection{Survey Results}

This section presents the results of the survey organised by the type of questions, first the demographics, then the Likert scales and finally, the open questions.\\

\subsubsection{Survey Respondent Demographics}
We received 83 answers in total, after cleaning the data, we ended up with 76 valid answers from Austria (1), Brazil (11), Ecuador (1), Spain (1), Germany (2), Ghana (1), Hungary (2), Italy (1), Mexico (17), Netherlands (1), Poland (1), South Korea (1), Sweden (33) and The United States (3).

Regarding pronouns, 57 respondents prefer ``he/him" pronouns, 14 prefer ``she/her," 4 opted for no pronouns, and 1 is comfortable with both ``she/her" and ``he/him".

The field experience of respondents varied as 4 (5.26\%) have less than 1 year of experience, 11 (14.47\%) have 1-2 years, 17 (22.37\%) have 2-5 years, 24 (31.58\%) have 5-10 years, and 20 (26.32\%) have over 10 years.\\

\begin{figure}
\centering
    \includegraphics[width=14cm]{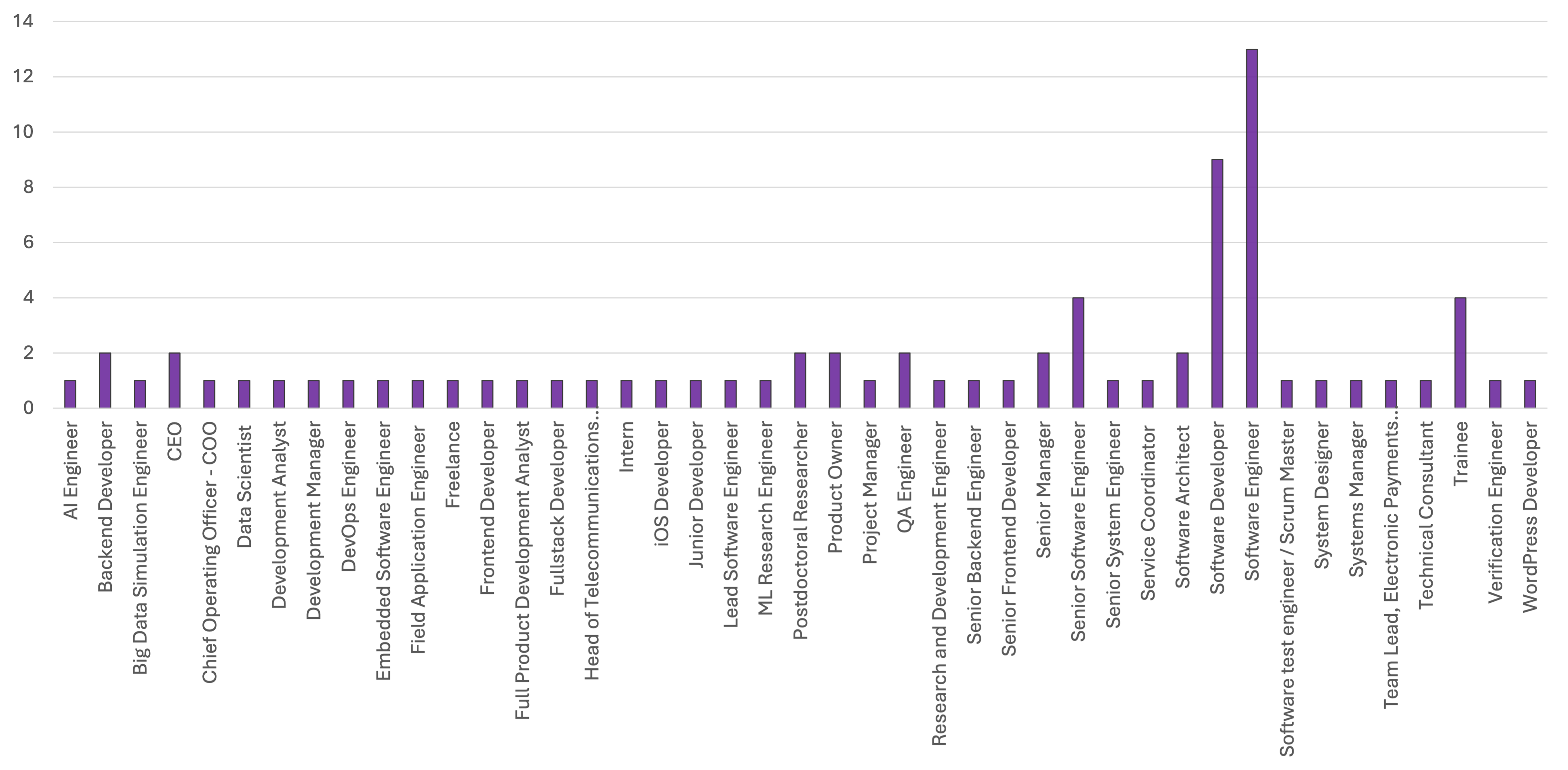}
    \caption{Respondent's positions.}
    \label{survey}
\end{figure}

\subsubsection{Likert Scale questions}

\begin{figure}
\centering
    \includegraphics[width=10cm]{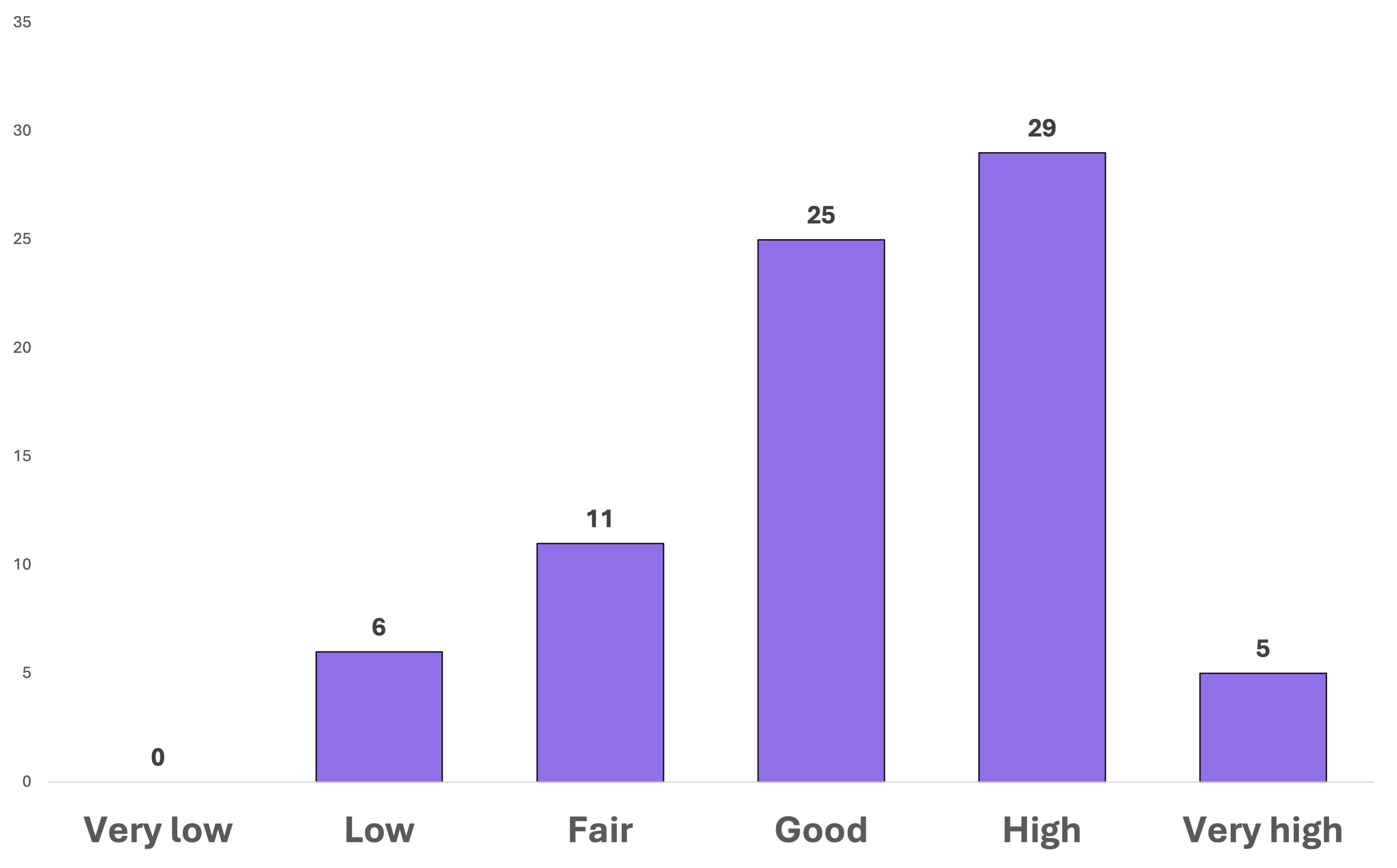}
    \caption{Overall well-being of respondents.}
    \label{overall}
\end{figure}

\begin{figure}
\centering
    \includegraphics[width=14cm]{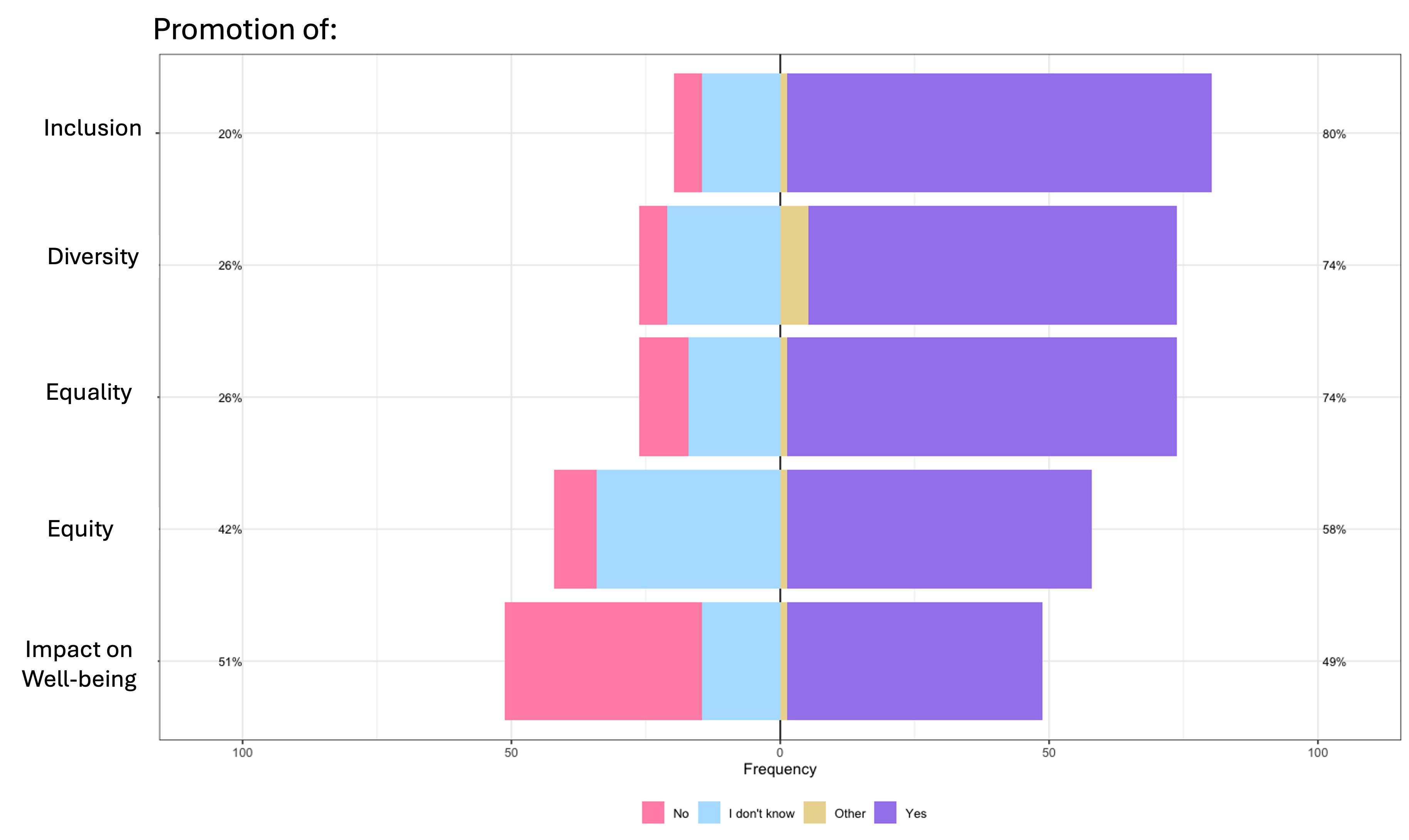}
    \caption{Answers about EEDI promotion in companies.}
    \label{eedi}
\end{figure}

The results from the Likert scale questions are presented in Figures \ref{overall}, \ref{survey} and \ref{eedi}. The overall well-being of our survey respondents is in general good, 38\% mentioned assessed it as high, 33\% as good, meanwhile only 8\% qualified as low and we did not get answers with very low.

Due to the nature of the questions we used different scales for each question. Figure~\ref{survey} shows the answers with its corresponding scale and percentages.

Our results revealed that most participants, 71\%, practice activities related to physical health and 51\% activities for mental health. When asked how often they face challenges with their teams, 67\% mentioned that occasionally and frequently and 33\% answered that they experienced negative impacts on their well-being due to colleagues or supervisors.

Most participants, 87\%, are satisfied with their work environment and 78\% with their compensation. Similarly, most of them feel respected (91\%) and heard (79\%). Further, participants perception of support in general (82\%), personal (74\%) and professional (64\%) was overall high. The quality of communication with their managers and peers was also mostly (83\%) rated towards the positive side. Finally, 48\% participants commented that their company's culture has an important influence on their well-being.

\begin{figure}[h!]
\centering
    \includegraphics[width=13cm]{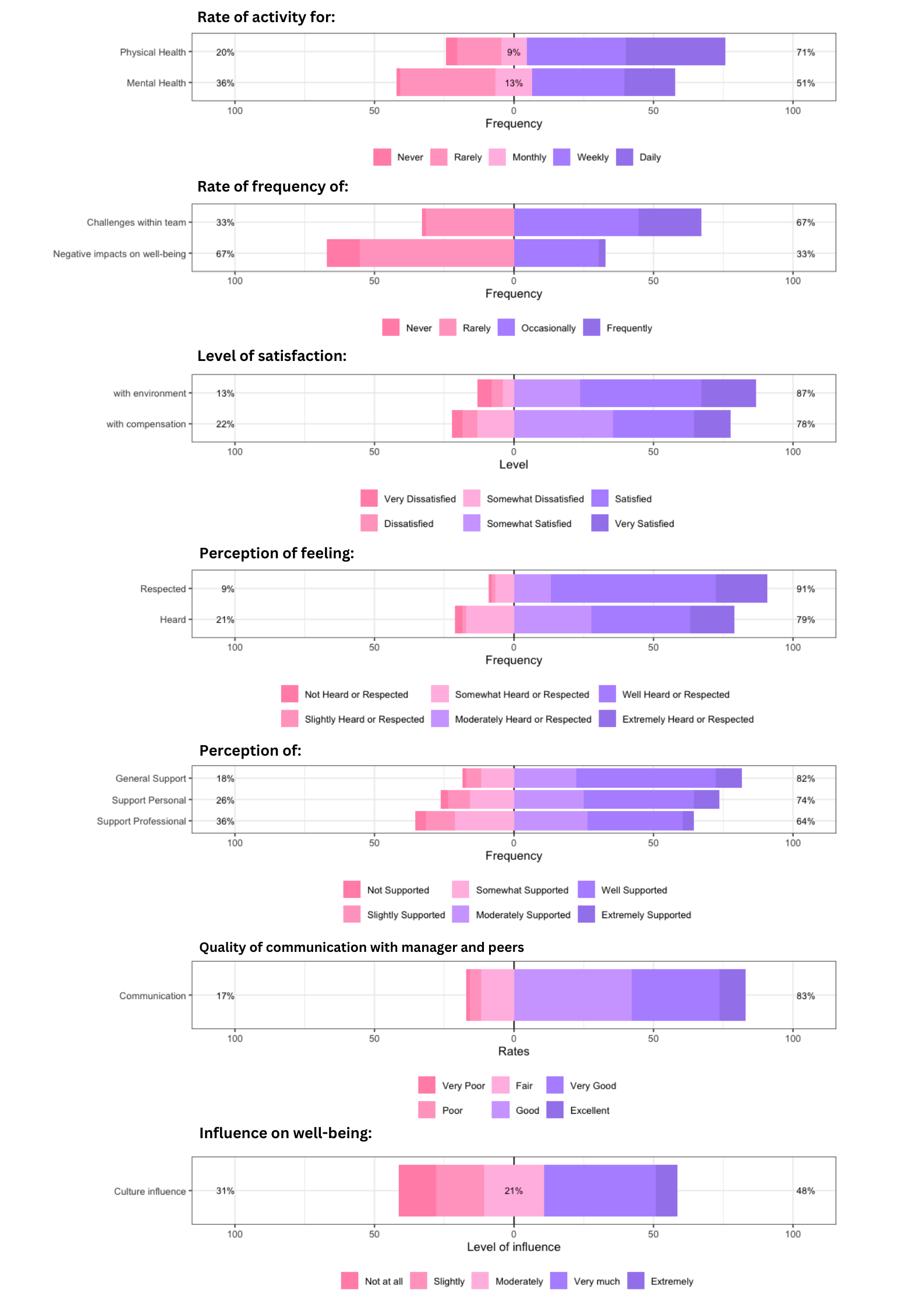}
    \caption{Diverging stacked bar chart showing results in percentage (\%) from the Likert scale survey questions (n = 76).}
    \label{survey}
\end{figure}

\begin{figure}
\centering
    \includegraphics[width=14cm]{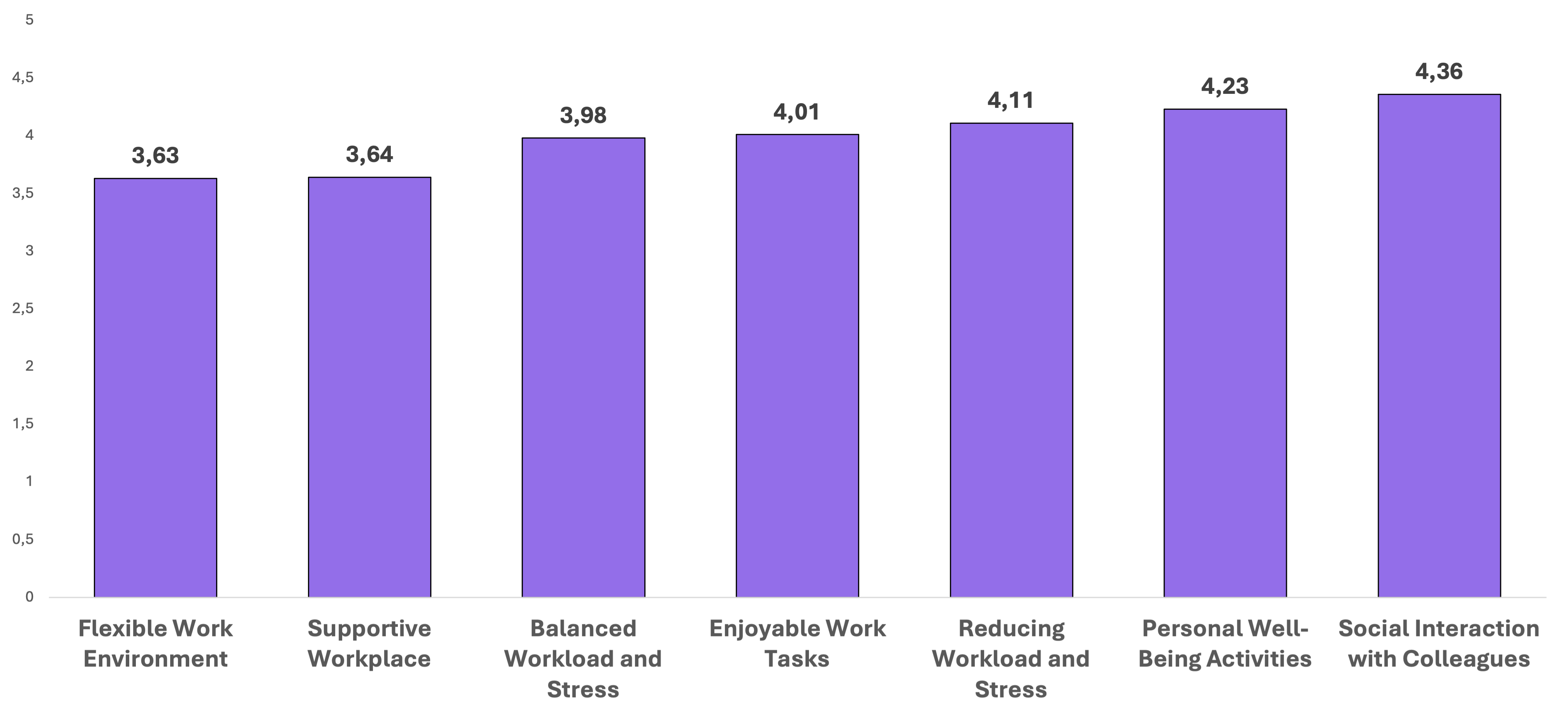}
    \caption{Factors contributing positively to respondents' well-being in the workplace showed per average.}
    \label{average}
\end{figure}

\begin{figure}
\centering
    \includegraphics[width=14cm]{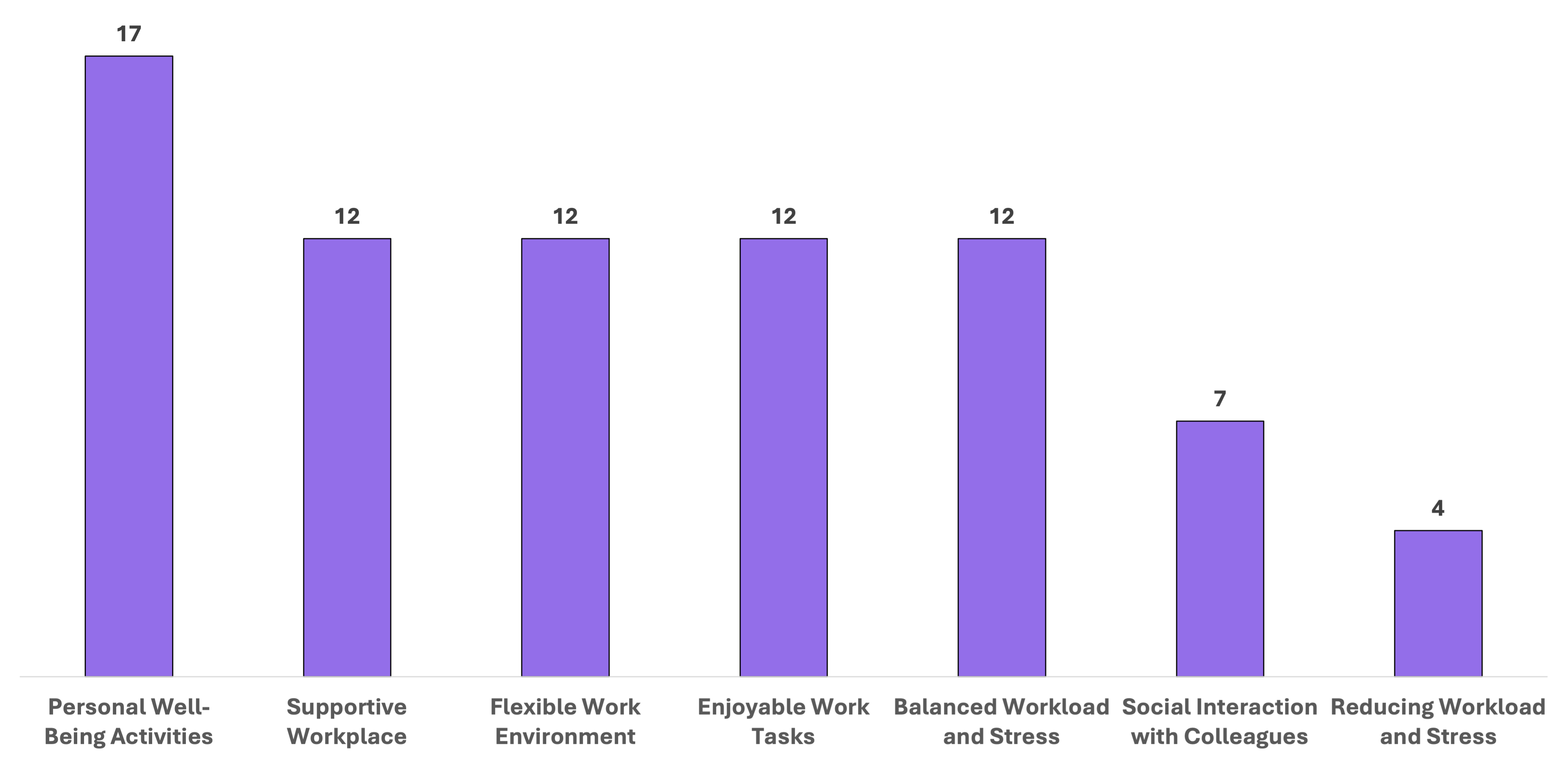}
    \caption{Factors contributing positively to respondents' well-being in the workplace. The graph shows the number of times each factor was chosen as the main factor contributing to their well-being.}
    \label{times}
\end{figure}

Figure~\ref{eedi} illustrates participants' views on whether their companies promote equality, equity, diversity, and inclusion (EEDI) and whether this promotion, or the lack thereof, affects their well-being. The majority of respondents indicated that their companies actively support EEDI initiatives and that these efforts positively impact their well-being.\\

Two questions were about the factors that contribute positively and negatively to the respondent's well-being (Q9 and Q10) and did not use a Likert scale. The questions were closed and participants had to choose specific answers, see Figure \ref{average}, \ref{times} for the result of Q9 and Table \ref{tab:challenges} for Q10. For the results in Figure \ref{average} and \ref{times}, participants had to rank from 1 to 7, a list of factors that contribute positively to their well-being. An average of the responses was made to obtain a visualisation, hence, the lowest average was the factor that was closest to 1 (most important), Flexible Work Environment. Furthermore, \ref{times} shows how many times each aspect was ranked as number 1, Personal Well-being Activities.

Regarding the factors or challenges they face in their workplace. Table \ref{tab:challenges} shows their answer in order of frequency. Personal life stress was chosen most times followed by a high workload. Excessive screen time and seasonal affective factors were mentioned the least.\\

\begin{table}[h]
  \caption{Challenges or obstacles affecting SE's well-being in the workplace}
  \centering
  \begin{tabular}{p{0.8\linewidth} p{0.05\linewidth}}
    \hline
    \textbf{Factor} & \textbf{\#} \\ \hline
    Personal life stress  & 47 \\[1ex]
    High workload  & 40 \\[1ex]
    Tight deadlines  & 35\\ 
    Challenges related to workplace communic. with managers  & 26\\ 
    Challenges related to workplace communication with peers  & 22 \\ 
     Pressure to keep up with rapidly changing technology  & 21 \\ 
     Seasonal affective factors, especially during winter  & 18 \\ 
     Excessive screen time  & 15 \\ \hline
  \end{tabular}
  \label{tab:challenges}
\end{table}


\subsubsection{Open Questions}
The open questions are presented in the coming subtitles. These questions were optional, hence, the amount of them was less in comparison with the Likerts questions.\\

\paragraph{Other factors that influence  respondents' well-being}

When asked about other factors besides the ones in Table \ref{tab:challenges} that negatively impact their well-being, participants mentioned that at an individual level, extended periods of isolation and issues related to migration are additional stressors. One respondent, who is a co-founder of a startup, feels a profound effect on their well-being based on the company's successes and failures. Concerning social interactions, participants mentioned peer pressure, the mental health issues of coworkers, boring relationships in the workplace, hostile work environment and communication issues, particularly with clients, as important factors. Regarding the company level, two participants cited traditional work environments with rigid schedules and resistance to hybrid or remote work as unnecessary and detrimental. An overwhelming workload, especially in areas outside one's expertise, micromanagement and the routine nature of work further contribute to a negative sense of well-being in the workplace.

On the opposite side, there were various answers regarding factors that positively influence respondents' workplace well-being besides the ones reported on~\ref{average}. One participant mentioned powerlifting as the only thing that works for them. Technological tools, specifically GPT-4, were also highlighted as beneficial, with one respondent expressing a positive impact from interacting with this AI. Food availability and quality play a crucial role; one participant commented that having a reasonably priced cafeteria and snack bar on-site allows them to not worry about meal preparation. Social interactions and recognition within the workplace were part of the answers, too; respondents cited the enjoyment of talking with friends at work and the positive effects of feeling listened to by management. Opportunities and recognition also emerged as key to enhancing well-being.\\

\paragraph{Influence of Company Culture on SE Well-Being}

The company culture significantly impacts SE's well-being, influencing various aspects such as work-life balance, inclusivity, engagement, support, management, mental and physical health, social interaction, motivation, and growth opportunities. Positive cultures that emphasise flexibility, support, inclusivity, meaningful work, and transparent management contribute to higher employee satisfaction and well-being. 

On the other hand, cultures that lack these elements can lead to stress, demotivation, and a negative impact on overall well-being. For instance, seven respondents mentioned that effective management and leadership are critical. The positive side includes transparency, collaborative environments, and a no-blame culture fostering safety and growth. The negative side includes poor management, lack of understanding from leaders, and hostile treatment towards employees.

Engagement is driven by meaningful work and alignment with personal values, based on five responses. Employees feel demotivated when their work seems pointless or disconnected from their values. Conversely, having a say in decision-making and understanding the company's goals enhances engagement. Similarly, five other participants agreed that the culture around work-life balance significantly affects their well-being. They appreciate flexible work hours, support for remote work, and the absence of micromanagement, all of which contribute to a comfortable and stress-free work environment. Further, a supportive environment, characterised by fun projects, the ability to change assignments, opportunities for continuous learning, a growth mindset, and group activities were recognised as crucial for five respondents. Social events and team bonding activities help employees build personal connections, which currently need to be improved in some companies. 

Regarding diversity, participants mentioned that a welcoming and inclusive culture, with representation of different people, positively impacts them. However, a lack of inclusivity, such as language barriers, can lead to fewer opportunities and feelings of exclusion. While some individuals feel unaffected by these initiatives, others report significant negative or positive impacts on their professional and personal lives. Respondents highlighted feelings of exclusion, frustration, and demotivation in environments that fail to promote EEDI. They emphasised the importance of feeling included, respected, and valued in the workplace. The mixed nature of the feedback suggests that while EEDI is a crucial factor for many, its importance varies widely depending on individual circumstances, work environments, and personal values.

Finally, three more answers talked about how having excessive meetings, high pressure, and a lack of understanding from management, as well as a focus on speed over quality, can lead to burnout and decreased motivation.\\

\paragraph{Respondents' Feedback on Workplace Relationships}

The answers collected indicate a general positive sentiment towards relationships with managers and peers. Communication, friendship, and supportive relationship dynamics were mentioned by participants. Six participants commented on positive relationship dynamics in their workplace. They mentioned having good and open-minded relationships with colleagues and managers, working well together, and having friendships between managers and team members; they also highlighted how these aspects positively affect the work environment and good camaraderie. Three participants commented on the crucial role of communication and tone in their workplace. Two more respondents mentioned the importance of addressing individual differences and providing support when needed. Finally, one mentions clients' behaviour and its impact on internal team dynamics and relationships.

Conversely, some participants mentioned several workplace challenges and areas for improvement. Issues such as perceived internal divisions, boring tasks, non-useful online meetings, and the mismatch between job demands and employee capabilities are notable. Additionally, one participant mentioned a need for greater transparency and acknowledgement. Further, one respondent mentioned that it is unnecessary to interact with coworkers outside of the workplace, while another commented that it is complicated to build bonds of trust with people who only listen to you for 15 minutes in the morning.  Finally, one last answer mentioned understanding and mitigating generational clashes as well as challenges to separate friendship and professional relationships as factors present in their current workplace.

Regarding maintaining team cohesion and effectiveness under stress, respondents indicated that effective communication, peer support, and strategic organisation are crucial for the team to achieve their goals. While many teams have developed robust strategies to cope with pressure, some struggle with disorganisation and over-reliance on individuals. Cultivating a supportive team environment and ensuring flexible, realistic planning appear to be critical factors in sustaining team well-being and productivity during challenging periods.\\

\paragraph{Recommendations given by participants to support their well-being}

In their recommendations, several participants mentioned that hybrid work should be allowed. Some commented that working from home has been great for their well-being. Additionally, they also recommended flexibility in schedules. 

One more participant commented to have workouts 2 -3 hours per week, walks, breaks to relax and more exercise activities. They also commented on giving complete or at least increased friskvårdsbidrag (Swedish health care allowance). Better salaries, bonuses for good work, more benefits, and considering effectiveness without putting pressure or micromanagement were also mentioned.

There were several points about managers, such as clarity in the tasks of managers and leaders, choosing qualified managers who know how to manage a team, prioritising personal coaching or mentoring over a traditional manager relationship, giving and implementing feedback, viewing employees as humans, and improving managers' training in human aspects to transmit knowledge and skills to their employees more effectively.

Some other recommendations were creating better workspaces designed to improve focus. Note that constant firing can decrease commitment, as employees may feel insecure about their job stability. It was also recommended to focus on increasing employee interaction, having informal meetings to discuss their challenges, and listening to their basic comfort needs. However, some other participants recommended reducing the number of meetings. They mentioned that addressing migrant issues can support their well-being, too. Additionally, employees appreciate having fruit baskets and plants in the office, which can contribute to a more pleasant and motivating workspace.

\paragraph{Final thoughts by participants on the personal well-being and the well-being of software engineers in general}

The final question was about anything participants wanted to add on their well-being or the well-being of software engineers in general; the answers highlighted various experiences, challenges, and recommendations.
Key well-being factors include maintaining a healthy work-life balance, accessing good work tools, fostering social interaction, and establishing personal routines. Respondents also valued environments that allow them to grow and feel connected to their work, and they recognised the importance of managing stress to maintain mental and physical health.

Furthermore, participants emphasised the importance of taking breaks, such as walks, to maintain well-being. Some expressed difficulty connecting with their employer and having difficulties finding motivation to perform well at work. Working from home was seen as beneficial for balancing work and family life, though it could blur the line between work and personal time. Others praised AI tools for easing their workload and improving productivity. Additionally, several respondents stressed the importance of physical exercise, proper ergonomics, and a good sleep routine. Some mentioned that software development can be lonely, and regular social interaction is necessary for well-being. Finally, they highlighted that motivation and enjoyment in work are crucial for maintaining overall satisfaction.

\section{Discussion} \label{sec:discussion}

Our findings reveal factors that influence software engineers' well-being across individual, team, peer, and organisational levels and indicate a varying significance of these factors.

In this section, we explore the main similarities and differences identified through our interviews and survey. We then compare our results with existing literature and models, highlighting both areas of alignment and points of divergence.

\subsection{Alignments in Quantitative and Qualitative Results}

\subsubsection{\textbf{Personal Practices}}

Interviewees and survey respondents reported frequently exercising three to five times weekly to support their well-being. This emphasis on exercise may serve as a coping mechanism, especially in a profession characterised by long hours, sedentary work, and high mental demands.

Survey results further indicate that personal well-being activities were identified 17 times as the most important factor influencing well-being, mirroring the insights from interviews and aligning with findings by Tsatsoulis and Fountoulakis \cite{tsatsoulis2006protective}. This reflects a strong sense of personal agency among software engineers, who actively engage in activities that help them decompress outside work. While personal efforts like exercise are undeniably valuable for stress management and maintaining well-being, broader factors -- such as long work hours, high cognitive demands, and a culture that often undervalues mental health -- also require attention.

\subsubsection{\textbf{Support from the Company and Peers}}
Participants in both the survey and interviews emphasised the importance of support from their company and peers. Survey results show that most respondents felt backed by company initiatives promoting a healthy work environment and employee well-being. Similarly, interviewees highlighted how peer support -- specifically in work-related matters -- fosters a positive and inclusive atmosphere, benefiting their mental health. These findings align with previous studies by Hirschle~\cite{hirschle2020stress} and Russo~\cite{russo2021predictors}, which also identified support as a key factor in mitigating the negative effects of stress and closely linked to increased productivity.

While employees report feeling supported by their company, the prevalence of stressors related to workload and time constraints indicates that these initiatives may not tackle deeper, systemic issues. The support provided seems not to extend to critical organisational changes such as improved project management, more realistic deadlines, greater workload flexibility, and enhanced support for hybrid work -- issues frequently raised by participants.

\subsubsection{\textbf{Work Environment: Trust, Physical Well-being, and Compensation}}
In the survey, participants expressed high satisfaction levels (87\% and 78\%) with their compensation and work environment. Meanwhile, in the interviews, participants emphasised trust as a key factor in their work environment. Interviewees highlighted that trust, particularly from management, was critical to their sense of well-being and ability to perform effectively. Trust was linked to a positive emotional state and practical aspects, such as flexibility in their roles and decision-making; this aligned with de Guerre et al. \cite{de2008structure}, who found trust as enabling conditions for mental health in organisations. Further, according to Syahreza et al. \cite{syahreza2017compensation}, compensation and work environment significantly impact employee satisfaction at work.

Satisfaction with financial compensation and physical work conditions is essential to maintaining a baseline level of employee contentment. However, these elements alone do not capture the full complexity of what makes a workplace genuinely supportive.

\subsubsection{\textbf{Equality, Equity, Diversity, and Inclusion (EEDI)}}
Several participants commented during the interview on company culture's openness to diversity, recognising efforts toward inclusion and equal opportunities. They acknowledged progress through diversity initiatives and team composition, while others pointed out persisting challenges, such as gender imbalances and glass ceilings.

In the survey, most respondents reported that their companies promote EEDI, which aligns with the interviewees. An important point to note is that most of the participants identified themselves with the pronoun ``him" (57/76). In contrast, the pronoun ``her" (14/76) and other pronouns (4/76) are a minority in our population. The minorities expressed stronger concerns and elaborated on their challenges, emphasising the need for a welcoming and inclusive culture.

There was a particular emphasis on how language barriers and lack of inclusivity often led to feelings of exclusion and missed opportunities, impacting directly on their well-being. De Souza and Gama~\cite{de2020diversity} obtained similar results when researching diversity and inclusion in IT companies.

While some respondents, particularly those from majority groups, were unaffected by EEDI initiatives, others reported both positive and negative impacts on their personal and professional lives. De Souza and Gama~\cite{de2020diversity} argue that the active involvement of majority groups in diversity efforts is crucial for driving change. However, achieving this can be difficult if these groups do not perceive the need for such change.


\subsubsection{\textbf{Personal Life Situations}}
Participants mentioned that their personal life situations can significantly impact their well-being, both positively and negatively, depending on the circumstances. Situations such as managing personal responsibilities, particularly family issues, added to work-related stress, however, supportive relationships and fulfilling personal activities were also highlighted as sources of positive well-being. Conditions like ADHD and challenges in balancing work and life were common themes. Our survey data confirmed this, with 47 respondents citing personal life stress as a significant factor affecting well-being. These results align with other studies identifying factors that contribute to poor mental well-being at work, such as Teevan et al. \cite{teevan2022microsoft} study finding integration of work and personal life, and by de Guerre et al. \cite{de2008structure} listing interpersonal conflicts as one of them. 

The struggle many employees face in balancing family responsibilities, personal challenges, and work demands likely stems from the rigidity of organisational structures. These structures typically lack the flexibility needed to accommodate diverse needs, such as flexible working hours or support for managing ADHD or family care responsibilities. As a result, employees are often expected to sustain high productivity while managing significant personal stressors without sufficient support. Although we did not analyse country-specific differences, it is clear that broader systems shape individual experiences differently (for instance, Sweden offers a better work-life balance) and offer a stark contrast. In organisations with rigid structures, the absence of flexible schedules, mental health resources, or accommodations for neurodivergent employees intensifies stress and diminishes employee engagement.

\subsubsection{\textbf{Workload and Time Constraints}}
Tied to the previous factor is the workload and time pressures. Participants reported that deadlines, customer demands, and poor allocation of responsibilities significantly impact their work experience. In the interviews, a lack of proper planning led to backlogs and increased stress, with employees feeling overwhelmed when client expectations exceeded the organisation's capacity. Survey responses also highlighted that high workload (40 respondents) and tight deadlines (35 respondents) were prominent sources of stress, which aligns with Scholarios and Marks \cite{scholarios2004work} and Teevan et al.~\cite{teevan2022microsoft} findings.

An overload of tasks, particularly in environments with poor planning, leads to a demotivated workforce, and without proper intervention, risks burnout and decreased long-term productivity.

\subsubsection{\textbf{Social Integration and Loneliness}}
Interviewees frequently discussed the difficulties of integrating socially within their teams and forming meaningful connections, especially in contexts where shyness or lack of social support networks created barriers to inclusion. Geographic factors (the country) and migration were mentioned as amplifying these feelings of isolation. Survey results support this, with participants citing social isolation as a significant stressor and naming issues like peer pressure and hostile workplace interactions. Other studies, such as D'Oliveira and Persico \cite {d2023workplace}, have reported on the effects of isolation on workplace well-being,  colleague and supervisor satisfaction, job satisfaction, and organisational commitment, aligning with our results.

The challenges of socialising and the resulting loneliness reflect individual characteristics like shyness and a workplace culture that may not facilitate inclusion or collaboration. This isolation is particularly pronounced for those who may be migrants or part of minority groups, as participants commented.

\subsubsection{\textbf{Tech Tools and Their Impact on Communication and Productivity}}
Both survey and interview participants mentioned frustrations with tech tools. These tools, such as Zoom or Teams, were seen as sometimes creating unnecessary meetings that could be replaced with emails, hindering productivity, which aligns with findings by \cite{nawrat2023tech}. Additionally, respondents complained about slow IT responses and tech tools that frequently crashed, leading to inefficiencies in communication and frustration.


\subsection{Contrasting Quantitative and Qualitative Results}

\subsubsection{\textbf{Influence of Social Interactions on Well-being}}

Interview participants highlighted positive social interactions and connections as crucial for emotional support and resilience in the workplace, directly influencing their well-being. They associated these connections with a sense of belonging, emotional support, and mental health, emphasising that positive workplace interactions create a more fulfilling and supportive environment.

In contrast, the survey results did not place as much emphasis on social interactions as a key factor in well-being (see Figures~\ref{average} and ~\ref{times}). While participants acknowledged social aspects—such as communication, friendship, and supportive relationships—these were framed as contributing factors rather than primary concerns. Other factors, such as personal well-being activities, flexible work environments, and overall workplace support, ranked higher in terms of impact on well-being.

The survey's lower prioritisation of social interactions may stem from participants focusing on more direct and measurable aspects of their work experience, such as workload, while viewing social dynamics as secondary. In contrast, interviews provided participants with more time to reflect on the broader range of factors affecting their well-being.

\subsubsection{\textbf{Recognition at Work}}

During the interviews, participants mentioned that feeling recognised and valued at work plays a significant role in their motivation and well-being, emphasising the importance of recognition. Meanwhile, in the survey, recognition was mentioned only as an ``other factor," with some respondents citing the positive effects of feeling listened to by management. However, it was not highlighted as a major contributor to well-being; it was grouped with minor factors.

\subsubsection{\textbf{Professional and Personal Growth Support from Companies}}

Regarding companies' professional and personal growth support, there were some differences in perceptions and opinions in the interview and survey. In interviews, participants expressed mixed feelings. Some felt there was a disconnect between their personal development goals and what the company offered, while others appreciated efforts like learning platforms and goal-setting opportunities. The survey respondents briefly mentioned growth opportunities as part of the overall company culture's impact on well-being. However, it was not a prominent focus compared to other factors like work-life balance and inclusivity. This aspect needs more research to draw solid conclusions; participants acknowledge the importance of growth opportunities and the need to align with individual career paths. Companies may need to tailor their initiatives to reach their employees' expectations and goals.

\subsubsection{\textbf{Company Policies and Practices}}
Interviewees valued company policies and well-being initiatives like wellness allowances and physical activity opportunities but expressed mixed feelings. Some commented to appreciate these efforts, while others felt insufficient and called for more comprehensive support.
The survey highlighted broader aspects of positive workplace cultures—emphasising flexibility, support, inclusivity, and meaningful work—as critical contributors to well-being. Hybrid work and work-life balance were frequently mentioned, but these were not mentioned in interviews. The difference in opinions can be due to the participants' contexts. All the interviews were done in Sweden, where hybrid work is already well established, while the survey covered different countries. Such countries may not have adapted hybrid work as Sweden has.

\renewcommand{\arraystretch}{1.5}
{\footnotesize
\begin{longtable}{p{0.18\textwidth} 
                  p{0.15\textwidth} 
                  p{0.15\textwidth} 
                  p{0.15\textwidth} 
                  p{0.25\textwidth}}
    \caption{Comparison to Other Theories and Models}\label{tab:factors} \\
    \toprule
    \textbf{Our Model} & 
    \textbf{Robinson’s Five Components of Well-being} & 
    \textbf{Seligman’s Five Pillars of Well-being} & 
    \textbf{Michaelson’s Pillars} & 
    \textbf{Added Value of Our Model} \\
    \midrule
    \endfirsthead

    \toprule
    \textbf{Our Model} & 
    \textbf{Robinson’s Five Components of Well-being} & 
    \textbf{Seligman’s Five Pillars of Well-being} & 
    \textbf{Michaelson’s Pillars} & 
    \textbf{Added Value of Our Model} \\
    \midrule
    \endhead

    \bottomrule
    \endfoot

    \bottomrule
    \endlastfoot

    Personal Conception of Well-being & - & - & - & Directly addresses personal interpretations of well-being, which the other models overlook \\
    \midrule
    \multicolumn{5}{l}{\textbf{Personal and Collaborative Factors}} \\
    \midrule
    Personal Practices & Physical well-being & Positive emotion & Emotional well-being, vitality, resilience, and self-esteem & Combines physical and emotional factors, acknowledging a broader scope of personal well-being practices \\\hline
    Influence of Social Interactions on Well-being & Social well-being & Relationships & - & Incorporates formal and informal social interactions inside and outside work, which are not fully considered in other models \\
    \midrule
    \multicolumn{5}{l}{\textbf{Support and Recognition}} \\
    \midrule
    Support from the Company and Peers on Well-being & Community well-being & - & - & Focuses on organisational and peer support, offering a more detailed look at company-level factors \\\hline
    Recognition at Work & - & Accomplishment & - & Directly addresses the importance of individual recognition at work on well-being, whereas others focus more on outcomes (e.g., accomplishment) \\\hline
    Professional and Personal Growth Support from Companies & Career well-being & Engagement & - & Emphasises the dual impact of personal and professional growth on well-being \\
    \midrule
    \multicolumn{5}{l}{\textbf{Work Environment and Culture}} \\
    \midrule
    Work Environment: Trust, Physical Well-being, and Compensation & Financial well-being & - & Positive functioning & Expands on workplace well-being by addressing trust and compensation in addition to physical and financial aspects \\
    Company Policies and Practices & - & - & - & Considers companies' well-being policies into the broader factors influencing well-being \\\hline
    Company Culture and Diversity & - & - & - & Considers companies' culture and efforts to achieve diversity as factors that contribute to well-being \\
    \midrule
    \multicolumn{5}{l}{\textbf{Challenges and Stressors}} \\
    \midrule
    Workload and Time Constraints & - & - & Positive functioning & Acknowledges the impact of workload and time pressures more explicitly than the other models \\\hline
    Social Integration and Loneliness & - & - & - & Stresses the importance of a person's sense of belonging and its influence on working life \\\hline
    Tech Tools and Their Impact on Communication and Productivity & - & - & - & Elaborates on how technology hinders and enhances work and its impact on well-being \\\hline
    Personal Life Situations & - & - & - & Acknowledges the positive and negative influence of personal life situations on working life \\
\end{longtable}
}


\renewcommand{\arraystretch}{1.5} 
\begin{table}[h!]
    \centering
    \caption{Policy Recommendations Based on Well-being Themes}
    {\footnotesize
    \begin{tabularx}{\textwidth}{p{0.25\textwidth} p{0.7\textwidth}}
        \toprule
        \textbf{Theme} & \textbf{Recommendation} \\
        \midrule
        \textbf{Personal Conception of Well-being} & Provide access to mental health resources, self-reflection exercises, and goal-setting programs that allow employees to understand their unique needs and preferences regarding well-being. Encourage and role model the use of such resources and activities to establish a culture of caring. \\
        \midrule
        \textbf{Personal and Collaborative Factors} & Create policies promoting individual well-being practices and positive interpersonal interactions at work using team-building activities and peer support networks. Encourage informal creative working spaces for ideas to flourish. \\
        \midrule
        \textbf{Support and Recognition} & Implement support systems that acknowledge and recognise both professional achievements. Similarly, strategies should be implemented to provide guidance and emotional support to ensure employees' well-being. Facilitate formal as well as informal mentoring and establish a visible role model culture. \\
        \midrule
        \textbf{Work Environment and Culture} & Develop policies that ensure a supportive and inclusive work environment, including trust, fair compensation, and diversity. Provide space and opportunity for the expression and exploration of local culture and diversity of culture if employees are from elsewhere. \\
        \midrule
        \textbf{Challenges and Stressors} & Create flexible work policies that address workload management (e.g., flexible work hours and hybrid work), social integration (e.g., virtual coffee breaks or social events for remote teams), and the use of technology (e.g., training in different tools). Provide parental leave or other care support to allow an employee to flourish while fulfilling family needs. \\
        \bottomrule
    \end{tabularx}}
    \label{tab:policy_recommendations}
\end{table}

\subsection{Comparison to Other Theories of Well-being Factors}

This section compares our model to other authors' theories. In Table \ref{tab:factors}, we align our findings with Robinson's five components of well-being, Seligman's five pillars of well-being, and Michaelson's pillars. Many of our themes align with these authors' proposals regarding the \textbf{multidimensional nature of well-being}. We placed each model in a separate column and listed the pillars or components that align with ours. Where there was no alignment, we indicated this with a `-'.  Consistent with these models, our study acknowledges that well-being is shaped by various factors, including emotional, psychological, social, and economic dimensions. Additionally, following Michaelson's work~\cite{michaelson2024national}, we advocate for integrating well-being into public policy, recognising that it reflects a broader understanding of the quality of life beyond economic growth alone.


While Wong et al.\cite{wong2023mental} study provides important insights into \textbf{internal experiences} and some \textbf{organisational factors}, we affirm that a more comprehensive approach is needed—one that considers and balances the external factors shaping well-being. Our model integrates a critical exploration of workplace dynamics, external pressures, and cross-cultural differences, offering a more nuanced understanding of how to support well-being in work environments. 

In contrast, Wong et al. primarily focus on \textbf{poor mental well-being at work}, addressing individual and organisational challenges, such as company culture, organisational policies, and personal coping strategies. While Wong et al.’s framework touches on external factors like \textbf{organisational culture and technologies} for mental well-being, it primarily focuses on internal self-reported experiences of well-being and the strategies software engineers use to manage it. This inward-looking focus, while important, \textbf{minimises the broader and more systemic external factors} that influence well-being, particularly those that are not under the direct control of individuals, such as workload demands, leadership dynamics, or cultural differences. Our approach \textbf{expands on Wong et al.’s results by emphasising the role of these external pressures} at every level—individual, team, and organisational. For instance, while Wong et al. acknowledge organisational challenges like company policies and culture, our research critically examines how specific external factors such as compensation, leadership practices, and structural job demands directly affect well-being. We argue that well-being is not just about how individuals or organisations manage mental health but also about how external factors shape the experience of well-being.

Additionally, Wong et al.'s study focuses on a \textbf{U.S. population}, which limits the generalisability of its conclusions. Our research expands the scope to include software engineers from various countries \textbf{worldwide}, allowing us to capture variations in organisational cultures, societal norms, and job structures that affect well-being.

\subsection{Policy Recommendations}

In our research work on well-being over the past five years, we observed that companies are unlikely to invest in well-being interventions that go beyond current policies. Recognising this, we have developed policy recommendations based on our research findings in order to enhance future policymaking.

Our recommendations are grounded in a rigorous analysis of the empirical data we collected through surveys and interviews for this study. By exploring well-being factors at individual, peer, managerial, and organisational levels, we identified key patterns, challenges, and opportunities related to the well-being of software engineers, and we reflected those findings in our guidelines.

One of the clearest and most consistent interpretations across our findings is the necessity for \textbf{flexible work policies that address workload management}. Several of our participants commented on their need for flexibility in their workplaces to ensure they take care of their needs outside work. Moreover, this recommendation stems from evidence indicating that flexible schedules can reduce stress and enhance productivity.

Table \ref{tab:policy_recommendations} shows these evidence-based recommendations on useful guidelines to 1) offer a roadmap for companies to effectively enhance the well-being of software engineers and 2) bridge the gap between research insights and practical policy. We aim to motivate organisations to implement measures that go beyond their current well-being frameworks, ultimately contributing to promoting a healthier and more resilient work environment and, hence, more resilient software engineers.

\subsection{Validity Threats}

This section outlines our study's possible threats to internal, external, and construct validity and the mitigation strategies we implemented. By identifying these threats and proposing mitigation strategies, the study aims to enhance the credibility of its conclusions about the factors influencing software engineers' well-being across different regions and cultural contexts.

\subsubsection{Internal validity} 

To affirm our internal validity and deal with selection bias, we targeted different sectors of software companies and engineers with different backgrounds. 

One more aspect we considered was the response bias. Participants may have given socially desirable answers during interviews, particularly when discussing sensitive topics like EDI or mental health. To encourage honest responses, we ensured anonymity and confidentiality during the interviews. Piloting the interview guide and survey helped us refine the questions' wording and tone to encourage more authentic answers. We also asked open-ended questions and used indirect questioning techniques to reduce pressure on participants to conform to perceived social norms.

\subsubsection{External validity}
We acknowledge that the internal validity can be compromised since our interviews were done only with software engineers working and living in Sweden, which may not represent the general population of software engineers. To mitigate this threat, we used purposive sampling to ensure diverse participants within Sweden (gender, ethnicity, company size.) to capture varied perspectives. Further, we targeted a broader sample with the survey. To ensure diverse representation, we aimed to recruit globally through various channels, including professional networks, social media, and industry groups and had our survey in three different languages.

Cultural and linguistic differences may influence perceived well-being, leading to inconsistent or incomparable results across regions. To mitigate this threat, we adapted the survey culturally in each language (English et al.) and worked with local experts to ensure that questions make sense in each context.

\subsubsection{Construct validity}

To ensure construct validity in the interviews, we defined concepts such as well-being, diversity, equality, equity and inclusion and gave examples for the interviews to make them explicit. Meanwhile, in the survey, we added definitions to the questions to ensure consistency in understanding across participants. Further, we ensured the translations aligned with the three languages we used. Continuing with the languages, we performed thorough back-translation of surveys and engaged local experts to ensure cultural nuances were considered. Pre-test translated surveys with small groups in each language to identify any problematic terms or misunderstandings.

Furthermore, we also tailored different scales to the questions in the survey in a way that measures each conception suitably.


\section{Conclusion}

To identify the main factors that influence software engineers' well-being, we conducted interviews in Sweden and ran a survey in three languages globally. We reported our main findings in this paper. 

Our study reports the main factors influencing well-being, such as \textbf{personal perception of well-being, personal and collaborative factors, support and recognition, work environment and culture, and challenges and stressors}. We confirmed the factors identified by research in other fields~\cite{seligman2011flourish,michaelson2024national} and offered unique contributions specific to the software engineering context.

First, we \textbf{strengthen the existing body of evidence} by analysing these factors in a field where high cognitive demands and constant technological evolution intensify their impact. Second, our model provides a \textbf{higher level of granularity}, identifying distinct stressors and the emotional toll they might have. We looked at these stressors at different levels, enabling deeper insights into how these factors manifest specifically within software engineering.

Third, our findings are \textbf{tailored to the software engineering population}, 
addressing nuances that general workplace studies often overlook. For instance, the critical importance of recognising individual contributions in team-based environments is particularly evident in this domain. Finally, we propose \textbf{a set of policy recommendations}, including flexible work structures and peer support networks, that directly address these challenges.

These contributions not only enhance understanding of well-being in this high-pressure field but also enable both practitioners and other researchers to develop interventions and support for these topic areas. 

Moreover, by systematically measuring various aspects of well-being, policymakers can make more informed decisions that improve overall quality of life, going beyond economic metrics that may not fully capture societal well-being and happiness.

Future work will include a more detailed analysis of country-specific differences. Additionally, we plan to gather input from managers on how they currently support software engineers' well-being and the outcomes of these efforts.


\section*{Acknowledgment}
We thank Ricardo Caldas for helping with the translation to Portuguese. We thank all participants for volunteering their time and personal experiences.

\bibliographystyle{ACM-Reference-Format}
\bibliography{bib.bib}

\vspace{12pt}

\end{document}